\documentclass[12pt]{aa}     
     
\usepackage{graphicx,amssymb,amsmath,times}     
\usepackage{longtable}     
\usepackage{bm}      
\usepackage{url}     
\usepackage[draft]{hyperref}     
\usepackage[varg]{txfonts}     
\usepackage[T1]{fontenc}     
\usepackage{natbib}     
\bibpunct{(}{)}{;}{a}{}{,} % to follow the A&A style     
\usepackage[normalem]{ulem}     
     
\def\simlt{\lower.5ex\hbox{\ltsima}}     
\def\simgt{\lower.5ex\hbox{\gtsima}}     
     
\def\wcen{$\omega$ Cen~}

\def\gtsim{\;\lower.6ex\hbox{$\sim$}\kern-6.7pt\raise.4ex\hbox{$>$}\;}     
\def\ltsim{\;\lower.6ex\hbox{$\sim$}\kern-6.9pt\raise.4ex\hbox{$<$}\;}

\titlerunning{Evolutionary and pulsation properties of Type II Cepheids}     
\authorrunning{Bono et al.}     
     
\begin{document}     
     
\title{Evolutionary and pulsation properties of Type II Cepheids}     
     
% ---Gruppo principale---     
\author{  
G.~Bono \inst{1,2}  
\and V.~F.~Braga \inst{2,3}  
\and G.~Fiorentino \inst{2}  
\and M.~Salaris\inst{4}  
\and A.~Pietrinferni \inst{5}  
\and M.~Castellani\inst{2}  
\and M.~Di Criscienzo\inst{2}  
\and M.~Fabrizio \inst{2,3}  
\and C.~E.~Mart{\'\i}nez-V{\'a}zquez \inst{6}  
\and M.~Monelli \inst{7}  
}  
\institute{  
Department of Physics, Universit\`a di Roma Tor Vergata, via della Ricerca Scientifica 1, 00133 Roma, Italy \\  
\and INAF-Osservatorio Astronomico di Roma, via Frascati 33, 00040 Monte Porzio Catone, Italy \\  
\and Space Science Data Center, via del Politecnico snc, 00133 Roma, Italy\\  
\and Astrophysics Research Institute, Liverpool John Moores University, IC2, Liverpool Science Park, 146 Brownlow Hill, Liverpool,L3 5RF, UK\\  
\and INAF-Osservatorio Astronomico d'Abruzzo, Via Mentore Maggini snc, Loc. Collurania, 64100 Teramo, Italy \\  
\and Cerro Tololo Inter-American Observatory, NSF's National Optical-Infrared Astronomy Research Laboratory, Casilla 603, La Serena, Chile \\  
\and Instituto de Astrof\'isica de Canarias, Calle Via Lactea s/n, E38205 La Laguna, Tenerife, Spain  
}

 \date{\centering Submitted \today\ / Received / Accepted }     
     
\abstract{We discuss the observed pulsation properties of Type II Cepheids (TIICs) in the   
Galaxy and in the Magellanic Clouds. We found that period (P) distributions, luminosity   
amplitudes and population ratios of the three different sub-groups (BL Herculis   
[BLH, P<5~days], W Virginis [WV, 5$\le$P<20~days], RV Tauri [RVT, P>20~days]) are   
quite similar in different stellar systems, suggesting a common evolutionary   
channel and a mild dependence on both metallicity and environment. We present a homogeneous   
theoretical framework based on Horizontal Branch (HB) evolutionary models, envisaging    
that TIICs are mainly old (t$\ge$10~Gyr), low-mass stars. The BLHs are   
predicted to be post early asymptotic giant branch (PEAGB) stars (double shell   
burning) on the verge of reaching their AGB track (first crossing of the instability strip),   
while WVs are a mix of PEAGB and post-AGB stars (hydrogen shell burning) moving from   
the cool to the hot side (second crossing) of the Hertzsprung-Russell Diagram.   
Thus suggesting that they are a single group of variable stars.    
The RVTs are predicted to be a mix of post-AGB stars along their second   
crossing (short-period tail) and thermally pulsing AGB stars (long-period tail)   
evolving towards their white dwarf cooling sequence.   
We also present several sets of synthetic HB models by assuming a bi-modal   
mass distribution along the HB. Theory suggests, in agreement with observations,   
that TIIC pulsation properties marginally depend on metallicity.   
Predicted period distributions and population ratios for BLHs agree quite well   
with observations, while those for WVs and RVTs are almost a factor of two   
smaller and larger than observed, respectively. Moreover, the predicted period distributions   
for WVs peak at periods shorter than observed, while   
those for RVTs display a long period tail not supported by observations.  
We investigate several avenues to explain these differences, but more detailed   
calculations are required to address these discrepancies.    
}     
     
\maketitle     
% These dates will be filled out by the publisher     
     
%%%%%%%%%%%%%%%%% BODY OF PAPER %%%%%%%%%%%%%%%%%%     
     
%_________________________________________________________________________     
\section{Introduction}\label{sec:intro}     
%_________________________________________________________________________     
     
The coupling between evolutionary and pulsation models      
has a long-standing tradition.    The reason is threefold.      
Firstly, pulsation models involve the stellar envelope, i.e. the portion of      
a stellar structure located between the region affected by nuclear burning      
and the stellar atmosphere.       
This means that they rely on the mass-luminosity relations and      
their temporal evolution predicted by evolutionary models.      
     
The comparison between pulsation predictions (periods, amplitudes, modal stability)      
and observations provide independent constraints on the input parameters (chemical      
composition) and on the micro- (opacity, equation of state) and the macro- (mass loss,      
convective transport) physics adopted to construct evolutionary and pulsation models.         
     
Finally, although the classification of variable stars is based on their pulsation      
properties and the morphology of the light curve,   
the evolutionary properties of variables stars provide fundamental      
insights on their progenitors and their different evolutionary channel(s).       
     
Seminal investigations concerning the coupling between evolutionary and pulsations      
models date back to the seventies, and have provided a firm evolutionary      
scenario for RR Lyraes (RRLs)\citep{iben70,tuggle73}, Classical   
Cepheids (CCs)\citep{kippenhahn67,fricke71,becker77}  and Mira \citep{vassiliadis93}   
variables.      
     
The improvement in the input physics (radiative [OP, \cite{iglesias96};      
OPAL, \cite{seaton07}] and molecular [\cite{ferguson05}]     
opacities) and in the treatment of the convective transport provided a new spin      
for both radial \citep{buchler88,chiosi93,kovacs93,bono94b,bono00b}     
and nonradial \citep{dziembowski99,guzik00}      
pulsators. The comparison between theory and observations experienced a significant      
step forward thanks to detailed grids of bolometric corrections and color-temperature      
transformations covering a broad range of chemical compositions, based on theoretical        
model atmospheres \citep{gustafsson75,kurucz79,castelli97,castelli03,gustafsson08}.     
     
In spite of the crucial role that type II Cepheids (TIICs\footnote{Note      
that we are suggesting to use the acronym TIICs, instead of T2Cs, to properly      
trace the key role that these variable stars  played in Baade's seminal 
discovery of Population~I and Population~II stars.}      
played in the cosmic distance scale     
\citep{baade58a,bono2016b} and the ongoing observational effort for the identification      
in the Galactic center \citep{matsunaga2013,braga2019b}, in the Galactic Bulge      
\citep{soszynski2011,soszynski2017}, in Galactic clusters \citep{matsunaga06} and in      
the Magellanic Clouds \citep{soszynski2018} the investigations concerning their      
evolutionary and pulsation properties lag when compared with other groups of      
radial variables. According to their pulsation properties (period distribution,      
shape of the light curve) TIICs can be classified into three different sub-groups     
\citep{soszynski08c}:      
BL Herculis (BLH) have periods longer than RRLs (P$\gtrsim$1 day) and shorter than   
five days, W Virginis (WVs) have      
periods between five and twenty days, while RV Tauri (RVTs) have periods longer      
than twenty days (see Fig.~1)\footnote{The RVTs should not be confused with   
Long Period Variables (LPV: Semiregular, Miras), because the latter group is located   
inside the so-called Mira instability strip. Although, TIICs and LPVs share similar   
evolutionary phases---double shell burning---the mean effective temperatures of LPVs   
are systematically cooler (see Fig.~1 in \citealt{martinezvazquez16b}).}.   
Although, the number of TIICs known in the literature increased      
by more than one order of magnitude in the last decade, detailed investigations      
concerning their evolutionary properties date back to the late seventies, early eighties.      
\cite{gingold74,gingold76,gingold85} provided exhaustive evolutionary calculations     
covering a broad range of stellar masses and chemical compositions. More recently,      
evolutionary and pulsation properties   
of TIICs were also investigated by \citet{bono97e}   
and \citet{dicriscienzo07},  but their analyses were   
limited to short-period TIICs (BLHs).       
     
The empirical scenario concerning TIICs has been enriched by several      
new interesting results. Based on  detailed multi-band investigation      
of Magellanic Cloud (MC) TIICs, \citet{groenewegenjurkovic2017} found that binary      
star evolution has to be taken into account for explaining their location in      
the Hertzsprung-Russell diagram (HRD). Note that binarity was originally invoked by      
\citet{soszynski08c} to explain the position in the Period-Wesenheit (PW) diagram      
of a sub-group of TIICs that, at fixed period, were systematically brighter      
than canonical TIICs.      
More recently, it has been suggested by \citet{iwanek2018} that MC TIICs      
might also have intermediate-age (0.5$\lesssim$t<10 Gyr) progenitors.   
Indeed, their 3D spatial distribution does not match that of RRLs      
(old, t$\ge$10 Gyr, stellar tracers), but it is half-way between RRLs      
and CCs (young---t$\lesssim$300 Myr---stellar tracers).      
     
Finally, it is worth mentioning that both the evolutionary channel producing RVTs      
and their pulsation properties are not clear yet. Indeed, observations     
suggest that this sub-group of TIICs includes both low-mass      
($\sim$0.5 $M_{\odot}$), and intermediate-mass ($\sim$1-2 $M_{\odot}$) stars.      
Moreover, the occurrence of alternating deep and shallow minima in the light      
curve and of infrared excess caused by circumstellar dust still lack a      
quantitative explanation.      
In this context it is worth mentioning that it is not even clear whether      
RVTs do follow the same Period-Luminosity (PL) relation of TIICs      
\citep{matsunaga06,bhardwaj17b,wallerstein2002,ripepi2015}.

The motivation for this investigation is threefold.     
Firstly, recent evolutionary grids of      
advanced evolutionary phases for low-mass stars cover a broad range in heavy      
elements and helium abundances. The same outcome applies to nonlinear, convective      
hydrodynamical models \citep{marconi07,marconi15,marconi2018}. Moreover,    
synthetic HB models are nowadays able to provide firm predictions concerning    
both evolutionary and pulsation models \citep{savino15}.     
     
On the observational side, long term optical (OGLE-IV, Catalina, PTF, PanStarrs, SDSS) and NIR (VVV,      
VMC, IRSF) surveys are providing a wealth of new identifications together with accurate      
pulsation properties (mean magnitudes, periods, luminosity amplitudes). Moreover,      
detailed photometric and spectroscopic      
investigations are opening a new path concerning the presence of TIICs in      
spectroscopic binaries, with the advantage to have an independent      
dynamical estimate of their actual stellar mass \citep{pilecki17}.     
     
Finally, current theoretical and empirical evidence indicate that the bulk of TIICs      
have old progenitors. This means that TIICs are excellent tracers of old stellar   
populations and thanks to their intrinsic brightness   
($\log{L}$/$\log{L_{\odot}} >$2) they can be identified   
in Local Group and in Local Volume galaxies.       
       
The structure of the paper is as follows. In \S~2 we summarize the     
pulsation properties (periods, amplitudes) of TIICs. In particular, we     
discuss the different classifications that have been suggested in the     
literature.   
Evolutionary properties of TIICs are dealt with in \S~3, 
where we address the role that pulsation and evolution      
play in explaining observed properties of TIICs. In \S~4 we discuss a      
set of synthetic HB models calculated to investigate the observed period      
distributions of TIICs, while in \S~5 we provide preliminary evolutionary      
calculations taking into account gravo-nuclear loops and He enhancement.       
In the last section we summarize our results and outlines possible future    
avenues for this project.     
Finally, the Appendix focuses on the metallicity distribution of field 
and cluster TIICs, and comparisons with the corresponding distribution 
for old stellar tracers (RRLs).   
     
%--------------------------------   FIG1  --------------------------------     
\begin{figure}[!htbp]     
\centering     
%\vspace{-100pt}     
\includegraphics[width=8.5cm]{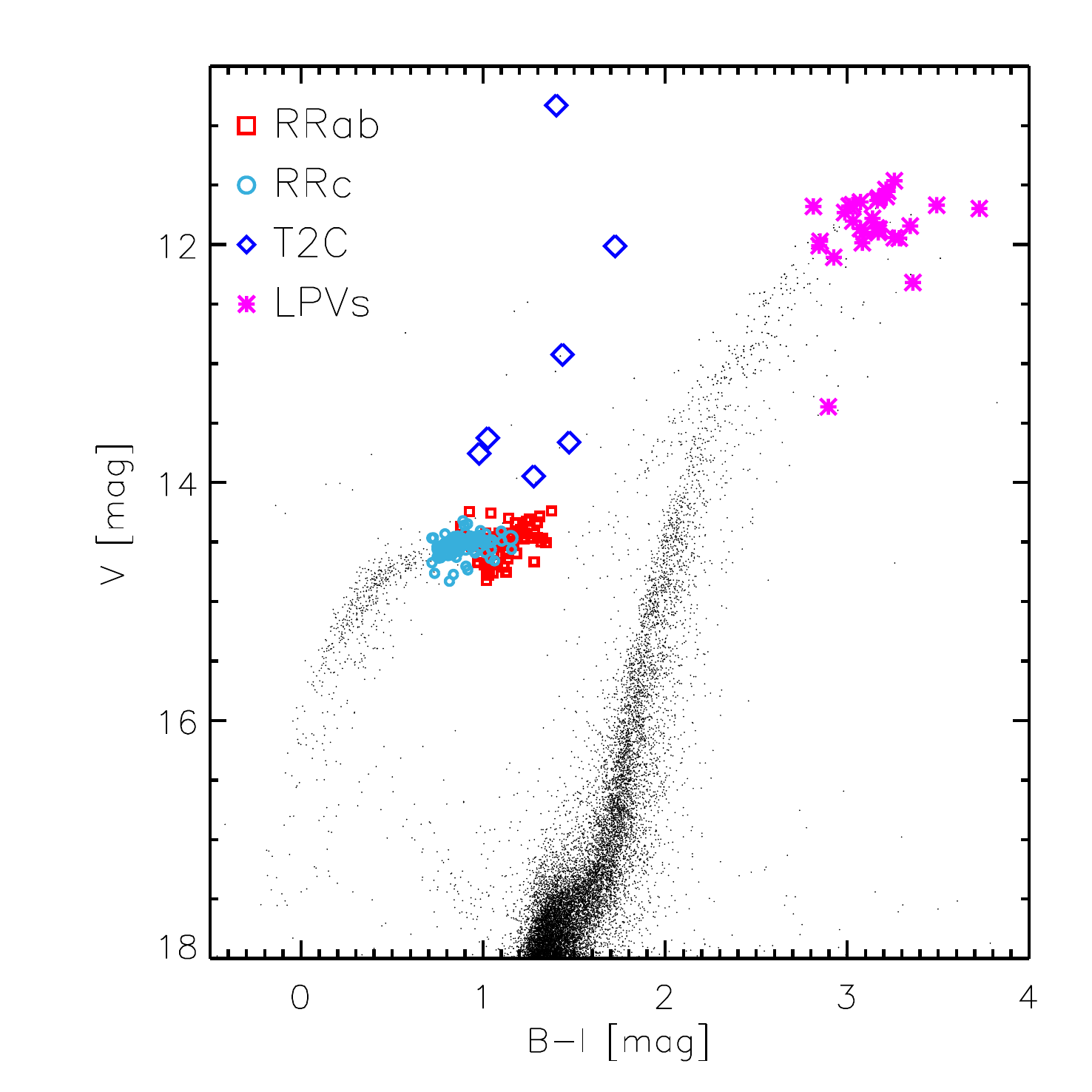}     
%%\vspace{-2cm}     
\caption[LoF entry]{     
B-I,V Optical Color-Magnitude Diagram of the Galactic globular $\omega$ Centauri     
\citep{braga16}. Black dots display cluster stars, selected according      
to radial distance and photometric error. 
The cyan circles and red squares display candidate cluster
first overtone (101 RRc, --0.6$\le\log P \le$--0.3)
and fundamental (85 RRab, --0.33$\le\log P \le$0.0) RRLs,
respectively \citep{braga16}. The blue diamonds denote TIICs 
(7, 0.05$\le\log P \le$1.5)), based on mean magnitudes derived in 
Braga et al. (2020, in preparation).
The magenta asterisks show a selection of cluster Long Period Variables
(LPVs, 1.58$\le\log P \le$2.72) identified by \citet{lebzelter2016},
but based on our own photometry.
}     
\label{fig_cmd}     
\end{figure}     
% temp_200328  
%-------------------------------------------------------------------------     
     
%_________________________________________________________________________     
\section{Observed pulsation properties of Type II Cepheids}\label{sec:pulsation}     
%_________________________________________________________________________     
     
%--------------------------------   FIG1  --------------------------------     
\begin{figure}[!htbp]     
\centering     
%\vspace{-100pt}     
\includegraphics[width=8.5cm, height=19cm]{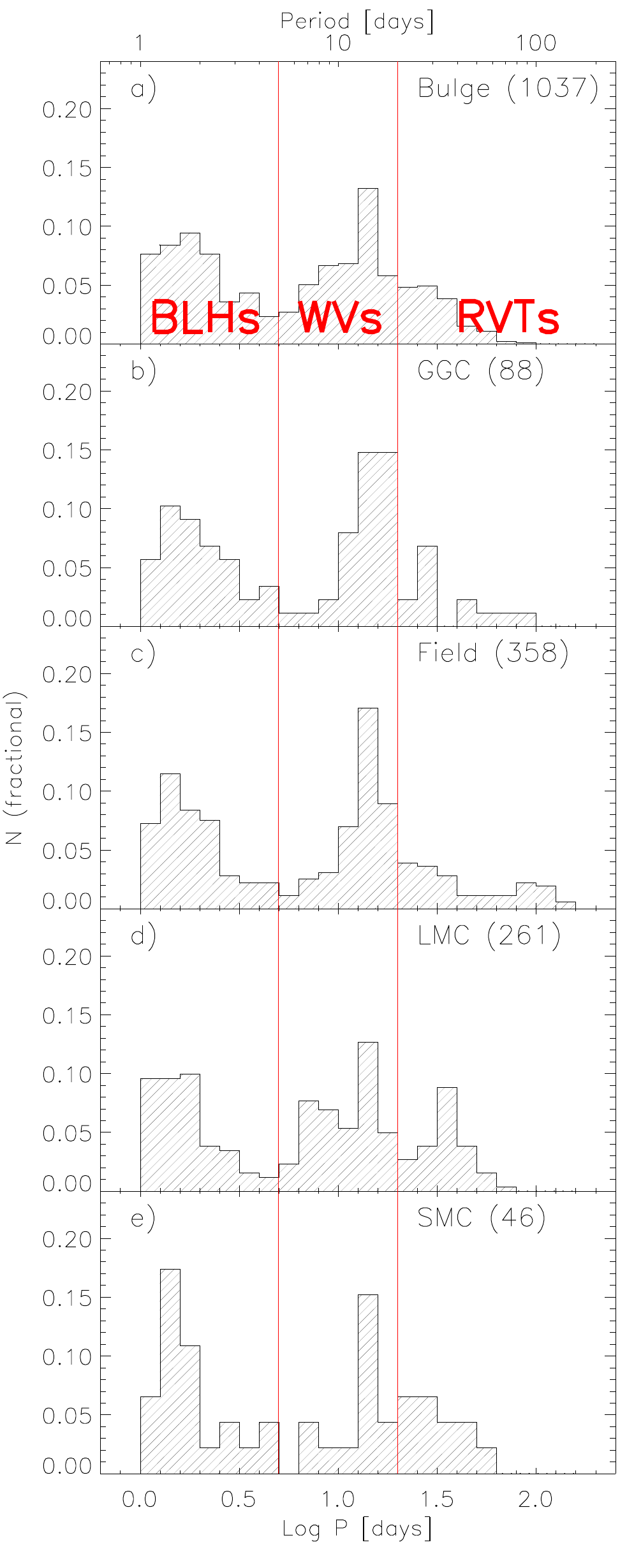}     
%\vspace{-0.5cm}     
\caption[LoF entry]{     
Panel a): Period distribution for TIICs in Galactic Bulge     
\citep[OGLE-IV,][]{soszynski2017}. The two red vertical lines display     
the period boundaries for the three sub-groups: BLHs (P$\le$5 days),     
WVs (5$< P\le$20 days) and RVTs (P$>$20 days)     
according to \citet{soszynski2011}. The total number of  TIICs     
and the number fractions of the three sub-groups are also labelled.     
Panel b): Same as Panel a), but for TIICs in Galactic globular clusters     
according to \citet{clement01,matsunaga06,pritzl03}.     
Panel c): Same as Panel a), but for TIICs in the Galactic field according     
to \citet{ripepi2019}.     
Panel d): Same as Panel a), but for TIICs in the Large Magellanic Cloud     
according to \citet{soszynski2018}.     
Panel e): Same as Panel a), but for TIICs in the Small Magellanic Cloud     
according to \citet{soszynski2018}.     
}       
\label{fig_period}     
\end{figure}     
     
%-------------------------------------------------------------------------     
     
To constrain the empirical pulsation properties of TIICs we      
collected data available in the literature for Galactic and MC   
variables. Concerning the Galactic ones we collected TIICs in the      
Bulge \citep[OGLE~IV, 1037,][]{soszynski2017},      
globular clusters (see Fig.~\ref{fig_period}) \citep{clement01,matsunaga06,pritzl03}      
and in the field \citep{ripepi2019}, while for the      
Magellanic Clouds ones we employ data provided by \citet{soszynski2018}.       
Figure~\ref{fig_period} shows the period distribution of both Galactic and      
Magellanic TIICs. They typically range from one to more than one hundred days.      
Moreover, they display two local minima for P$\sim$5 and      
P$\sim$20 days. The former one was adopted for separating BLHs      
from WVs and the current data are, indeed, suggesting that it ranges from      
about four days in the Bulge to about six days in the Galactic      
field. The latter one (P$\sim$20 days) was adopted for separating WVs from      
RVTs, and the current data are suggesting that it shows up      
as a shoulder in the period distribution of Bulge and field TIICs and      
as a local minimum in GCs and in MC TIICs.        
     
Data plotted in this figure display several interesting features worth      
being discussed in more detail.      
     
First of all, the relative fractions of BLHs and WVs are the same 
within 1$\sigma$ (Poisson uncertainty) when moving from Galactic (panels a, b, c) to Magellanic TIICs  
(panels d, e). The values are on the order of 40\%, suggesting a common evolutionary  
channel and a mild dependence on the initial metal content.  
Regarding the fraction of WV in the Small Magellanic  
Cloud (SMC), the difference is still within 1$\sigma$, but the Poisson   
uncertainty is twice as large as for the LMC, because the total  
number of TIICs in this SMC is still limited. An increase by a  
factor 5-6 in the sample size is required to reach firmer conclusions.  
The relative fraction of RVTs in the same stellar systems  
ranges from 15\% to 25\%. The differences are on the order of 1$\sigma$,  
but once again, the SMC sample size is limited.  
Moreover, the period distribution changes significantly when moving  
from the Bulge to the Galactic field and to the MCs (see Fig.~\ref{fig_period}).  
This points to an observational bias due      
to the possible misclassification of the alternate deep and shallow minima      
characterizing this group of variable stars \citep{percy1991}.       
  
Regarding WVs, their period distribution appears to be peaked at      
$\log P\sim$1.1-1.2, quite uniform in the Bulge and in the LMC,   
while the distribution is skewed toward longer periods in GCs and   
in the Galactic field.      

% Ref. 1 
%
Our referee suggested to quantify better the difference in the period
distribution of TIICs when moving from the Galaxy to the MCs. We performed
a nonparametric, two-tailed Kolmogorov-Smirnov test by assuming the period
distribution of the Bulge TIICs as the reference distribution.
We found that on average the probability of the null hypothesis --that the
two distributions come from the same population-- is on the order of
$\sim$60\% for the pairs Bulge--SMC and Bulge--Halo, while it increases to
73\% for the pair Bulge--GGCs, and to 92\% for the pair Bulge--LMC.
The variation of the probability of the null hypothesis by almost 50\% when
moving from the pairs Bulge--SMC/Bulge--Halo to the pair Bulge--LMC further
supports the broad variety of TIICs in different environments.

%--------------------------------   FIG2  --------------------------------     
     
\begin{figure}[!htbp]     
\centering     
\includegraphics[width=8.5cm, height=19cm]{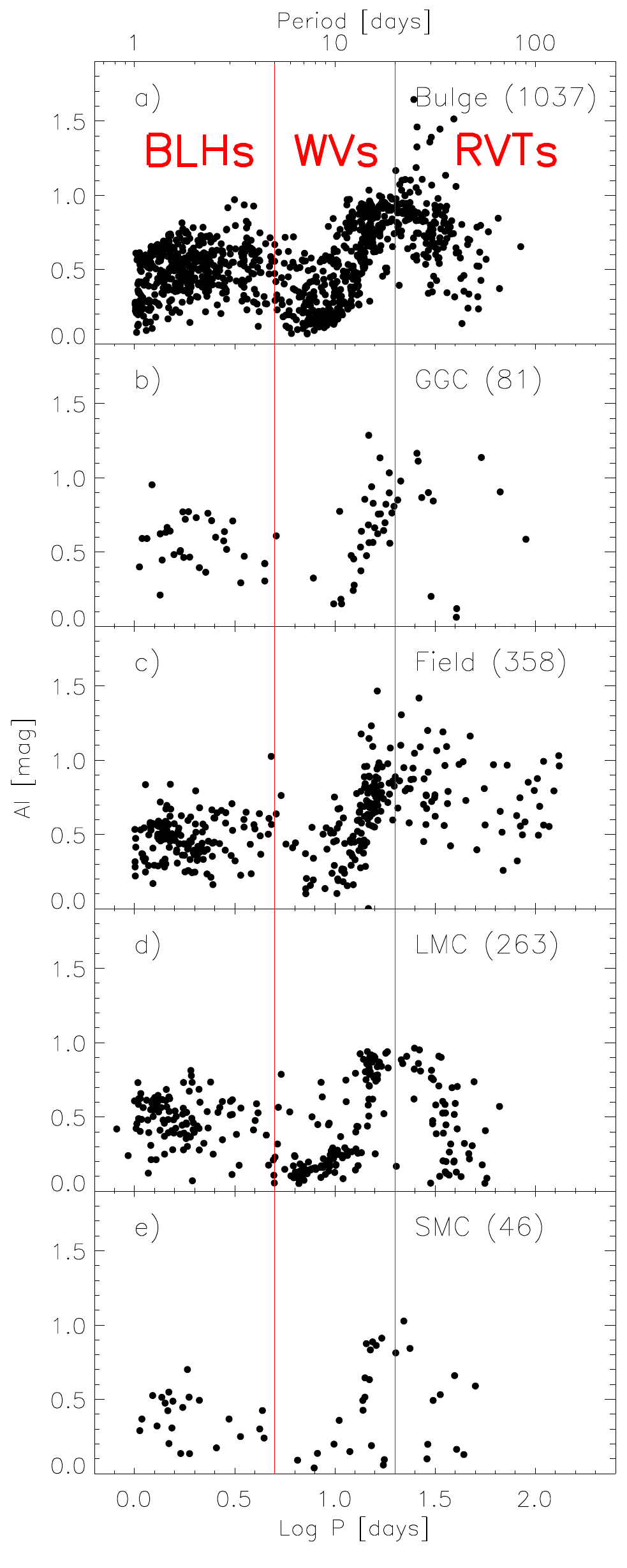}      
%\vspace{-0.5cm}     
\caption[LoF entry]{     
I-band amplitude versus logarithmic period (Bailey diagram) for Bulge     
TIICs (panel a), GC TIICs (panel, b), Galactic field TIICs (panel c),     
Large Magellanic Cloud TIICs (panel d) and Small Magellanic Cloud TIICs     
(panel e). The red vertical lines are the same as in Fig.~\ref{fig_period}     
(see text for more details).     
}     
\label{fig-bailey}     
\end{figure}     
     
%-------------------------------------------------------------------------     
     
The difference in the period distribution among BLHs, WVs and RVTs is       
fully supported by the Bailey diagram, the $I$-band luminosity      
amplitudes versus logarithmic period shown in (Fig.~\ref{fig-bailey}).      
The $I$-band amplitudes for Bulge and Magellanic TIICs were      
provided by \citet{soszynski2017,soszynski2018}. For GGCs, we 
adopted the $B$-, $V$- and $K$-band luminosity amplitudes of cluster TIICs      
from \citet{clement01} and transformed them into the $I$-band employing the amplitude      
ratios derived by Braga et al. (2020, in preparation). The latter
ratios were derived using OGLE $VI$- and VVV $K_s$-band 
light curves of Bulge TIICs and $BV$-band      
light curves of TIICs in GGCs. The Bailey diagram shows that   
WVs attain a well defined minimum at  P$\sim$8 days,   
with a steady increase toward longer periods. The trend      
for RVTs is far from being homogeneous, because the maximum around      
twenty days is broad. Moreover, RVTs in the Bulge and in the LMC display      
a steady decrease toward longer periods and a well defined cut-off at periods      
longer than 60 days. On the other hand, RVTs in the Galactic field      
approach 200 days and display at fixed period a broad range in      
luminosity amplitudes.        
     
The current partition of TIICs into three sub-groups follows the classification      
suggested by \citet{soszynski08c,soszynski2011}.      
They also suggested a new group of TIICs, the      
peculiar WVs (pWVs) which have peculiar light curves. Moreover the pWVs are,      
at fixed period, brighter than typical TIICs.      
These are the reasons why the pWVs are not included in this   
analysis.  
Data plotted in Fig.~2 and in Fig.~3, based on observables that are      
independent of distance and reddening, shows that the current classification      
is plausible. However, the boundaries of the different sub-groups      
might be different in different stellar systems, suggesting that the      
environment and the mean metallicity might globally affect their   
properties.   
  
The possible dependence on the metallicity requires a more detailed discussion.      
We still lack spectroscopic measurements of Bulge TIICs, so we assume that their      
metallicity distribution is either similar to the one of Bulge RRLs as measured by \citet{walker1991a},       
suggesting a mean [Fe/H]=$-$1.0  with a 0.16~dex standard deviation,       
or similar to Bulge red giant stars, with average [Fe/H]=$-$1.5 and a standard deviation equal to 0.5~dex      
\citep{rich2012,zoccali2017}.     
The metallicity distribution of TIICs in GCs and in the Galactic field     
ranges from [Fe/H]$\sim -$2.4 to slightly super solar [Fe/H]       
(see Appendix~\ref{sec:metallicity}).        
Concerning LMC TIICs, we can follow two different paths.      
According to \citet{gratton04a} the metallicity of LMC RRLs based on low-resolution      
spectra range from [Fe/H]=--2.1 to  [Fe/H]=--0.3, but only a few stars are more      
metal-rich than [Fe/H]=--1, indeed, the mean metallicity for 98 RRLs is      
[Fe/H]=--1.48$\pm0.03 \pm0.06$. This metallicity range is also supported by recent      
investigations concerning the mean metallicity of LMC globular clusters.   
Using homogeneous Stroemgren photometry,     
\citet{piatti2018} found, in agreement with      
spectroscopic measurements, that the metallicity of the ten LMC globulars ranges from          
$-$2.1 dex (NGC~1841) to $-$1.1 dex (ESO121-SC3). We still lack direct measurements of the      
metallicity distribution of truly old SMC stellar tracers, however the metallicity of      
NGC~121, the only SMC globular cluster, is [Fe/H]=$-$1.28,      
according to high resolution spectroscopy \citep[][]{dalessandro2016}.   
Table~\ref{tab:cluster_met}  gives either the mean metallicity or      
the metallicity range of the stellar systems     
included in Fig.~\ref{fig:metallicity}. The above      
evidence shows that the TIICs plotted in Fig.~\ref{fig:metallicity} 
cover roughly three dex in metal  abundance, but the population ratios 
appear to be, within the errors, quite similar.        
  
%%%%%%%%%%%%%%%%%%%%%%%%%%%%%%%%%%%%%%%%%%%%%%%%%%%%%%%%%%%%%%%%%%%%%%%%5  
% 			Table 1 
%%%%%%%%%%%%%%%%%%%%%%%%%%%%%%%%%%%%%%%%%%%%%%%%%%%%%%%%%%%%%%%%%%%%%%%%5  
\begin{table}  
\scriptsize  
\caption{Population ratios for TIICs in different Galactic components and   
in the Magellanic Clouds.}  
\label{tab:countratio}  
\centering  
\begin{tabular}{lccccc}  
\hline  
\hline  
Type & \multicolumn{5}{c}{Environment} \\  
&  Halo & Bulge & GGC &  LMC & SMC \\  
\hline  
BLH    & 0.408$\pm$0.040 & 0.431$\pm$0.024 & 0.432$\pm$0.084 & 0.379$\pm$0.045 & 0.435$\pm$0.116 \\  
WV     & 0.439$\pm$0.042 & 0.405$\pm$0.023 & 0.420$\pm$0.082 & 0.410$\pm$0.047 & 0.326$\pm$0.097 \\  
RVT    & 0.154$\pm$0.022 & 0.164$\pm$0.014 & 0.148$\pm$0.044 & 0.211$\pm$0.031 & 0.239$\pm$0.080 \\  
\hline  
\end{tabular}  
\end{table}  
%%%%%%%%%%%%%%%%%%%%%%%%%%%%%%%%%%%%%%%%%%%%%%%%%%%%%%%%%%%%%%%%%%%%%%%%5  
  
%_______________________________________________________________________________     
\section{Properties of Type II Cepheids}\label{sec:evolution}     
%_______________________________________________________________________________     
     
TIICs have been the cross-road of several theoretical and empirical      
investigations, however, their evolutionary status is far from being well established.      
A first analysis of the evolutionary properties of TIICs was      
provided over 40 years ago by \cite{gingold74,gingold76,gingold85}.      
He recognized that a significant fraction of hot (blue) HB stars evolve       
off the Zero-Age-Horizontal-Branch (ZAHB) from the blue (hot)      
to the red (cool) region of the CMD.     
In the approach to their AGB track these stars are in a double shell (hydrogen and helium) burning phase      
\citep{salaris05} and cross the instability strip at luminosities systematically      
brighter than typical RRLs. The difference in luminosity and the smaller mass     
compared to RRLs induce an increase in the pulsation period of TIICs      
when compared with RRLs. Typically, the two classes are separated by  
a period threshold at one day. This separation   is supported by a well 
defined minimum of the period distribution when moving from RRLs to TIICs, 
but the physical mechanism(s) causing this minimum are not yet clear, and   
the exact transition between RRLs and TIICs has not been established yet 
(Braga et al. in preparation).   
     
The quoted calculations suggested also that blue HB stars after the first crossing of      
the instability strip experience a co-called \lq{blue nose\rq} (then dubbed \lq{Gingold's nose\rq}),     
e.g. a blue-loop in the CMD, that causes two further crossings of the instability strip before     
the tracks reach the AGB.      
These three consecutive excursions were associated to the interplay between the helium      
and hydrogen burning shells. After core helium exhaustion, HB models with      
massive enough envelopes evolve redwards in the CMD. The subsequent shell helium ignition causes      
a further expansion of the envelope, and in turn, a decrease in the efficiency      
of the shell hydrogen burning, which causes a temporary contraction of the envelope,   
and a blueward evolution     
in the CMD. Once shell hydrogen burning increases again its energy production,       
these models move back towards the AGB track.       
Finally, these models would eventually experience a fourth blueward      
crossing of the instability strip before approaching their white dwarf (WD)      
cooling sequence (see Fig.~1 in \cite{gingold85} and Fig.~2 in \cite{maas2007}).       
During their final crossing of   the instability strip, models (in the post-AGB phase),     
are only supported by a vanishing shell H-burning.      
     
Basic arguments based on their evolutionary status and the use of the pulsation      
relation available at that time \citep{vanalbada73} allowed Gingold to      
associate the first three crossings (including the \lq{Gingold's nose\rq}) to      
BLHs and the fourth one to the WVs variables.       
     
These early analyses, however, lacked qualitative estimates of the time spent      
inside the instability strip during the different crossings, and in particular,      
the period distributions associated to the different crossings. Moreover and even      
more importantly, dating back to more than 25 years ago, HB evolutionary models       
based on updated physics inputs,      
\citep{lee90,castellani91,dorman93,brown00,pietrinferni06a,dotter08,vandenberg13}     
do not show the \lq{Gingold's nose\rq.

%^^^^^^^^^^^^^^^^^^^^^^^^^^^^^^^^^^^^^^^^^^^^^^^^^^^^^^^^^^^^^^^^^^^^^^^^^^^^^^^     
% Ref. 3a
\subsection{Pulsation properties of Type II Cepheids}\label{sec:3.1}     
%^^^^^^^^^^^^^^^^^^^^^^^^^^^^^^^^^^^^^^^^^^^^^^^^^^^^^^^^^^^^^^^^^^^^^^^^^^^^^^^     
     
We have already discussed in sections 1 and 2 the change in the topology of the    
instability strip, i.e. the regions in the HRD with stable modes of pulsations   
for RRLs and TIICs.   
We adopted the theoretical instability strip for RRLs recently provided by      
\citet{marconi15}. Note that the blue and red boundaries have been       
extrapolated to higher luminosities to cover the luminosity range typical      
of TIICs. The reason why we decided to follow this approach is twofold.       
First of all, the instability strips provided by \citet{marconi15} cover a broad range in      
metal abundances, stellar masses and luminosities typical of RRLs. This means that      
they cover the entire range of RRLs and the short-period range for TIICs.   
Secondly, this is an exploratory investigation, to trace       
the global evolutionary and pulsation properties of TIICs. More detailed      
calculations concerning the topology of the instability strip covering   
both short- and long-period range of TIICs will be provided in a   
forthcoming investigation.   
   
%^^^^^^^^^^^^^^^^^^^^^^^^^^^^^^^^^^^^^^^^^^^^^^^^^^^^^^^^^^^^^^^^^^^^^^^^^^^^^^^     
\begin{figure}[!htbp]     
\centering     
\includegraphics[width=8.5cm,height=10cm]{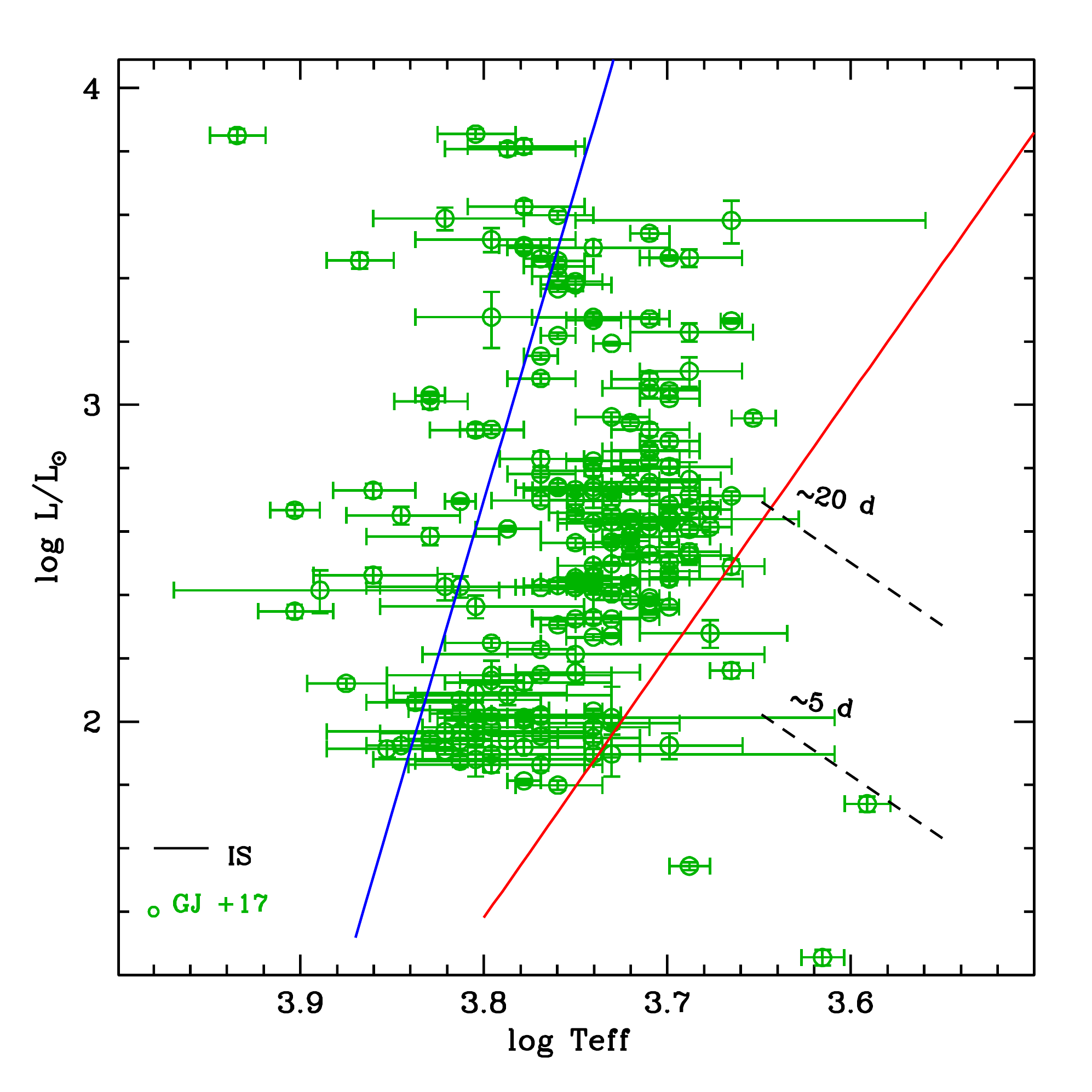}     
\caption{Hertzsprung Russell Diagram (HRD) comparing the      
predicted (blue and red solid lines) edges of the instability strip and      
observations for Magellanic TIICs by      
\citet{groenewegenjurkovic2017,groenewegenjurkovic2017b}.     
The horizontal error bars display the uncertainty in effective temperature.     
The black dashed lines display two iso-periodic lines for five and twenty   
days, respectively.     
}     
\label{dataoss}     
\end{figure}     
%^^^^^^^^^^^^^^^^^^^^^^^^^^^^^^^^^^^^^^^^^^^^^^^^^^^^^^^^^^^^^^^^^^^^^^^^^^^^^^^     
     
To validate the predicted boundaries of the instability strip, Fig.~\ref{dataoss} shows      
the comparison in the HR diagram with observations for Magellanic TIICs     
recently provided by \citet{groenewegenjurkovic2017,groenewegenjurkovic2017b}.      
Owing to the lack of spectroscopic estimates of the metal abundances,      
we assumed a metallicity range between $\sim -$2.2 dex and $-$0.7 dex      
(see Appendix~\ref{sec:metallicity}). To compare theory and observations we derive    
mean boundaries of the instability strip      
by taking into account models with metal mass fractions ranging from Z=0.0001 to Z=0.01.      
Data plotted in this figure shows that theory and observations agree quite well,    
even though, the predicted instability strips have been computed   with a step in    
effective temperature of 100 K and they have been extrapolated to brighter    
luminosities.     
     
Moreover, observations display a large spread in effective temperature when      
moving from fainter to brighter TIICs, but the number of objects      
hotter than the blue edge is, within the errors, limited. Note that      
the current predictions for the blue edge of the instability strip are      
quite solid, since it is minimally affected by uncertainties in the      
treatment of the convective transport \citep{baker1979}. It would be      
interesting to extend the comparison between theory and observations      
into optical and near-infrared color magnitude diagrams, to      
further constrain the plausibility of the current predictions.       
It is worth mentioning that the predicted strip has a width in temperature   
ranging from $\approx$1200~K in the period range of BLHs, to $\approx$1600~K   
in the period range typical of WVs. These estimates agree quite well   
with similar estimates for the width in temperature of the RRL instability   
strip \citep{bono94b,tammann03}. A wider instability   
strip for TIICs would cause a dispersion in luminosity at fixed period, 
significantly larger than currently observed.   

% Ref. 4
The anonymous referee suggested a more quantitative assessment of the
width in temperature of the instability strip. We selected RRLs, TIICs
and CCs in the LMC, because they are complete samples 
\citep[OGLEIV,][]{udalski2015} and used the standard deviation of the
PW($I$,$V-I$) relation as a proxy for the width in temperature of the instability
strip \citep{bono99b,bono08,madore17,riess2019}.
% Ref. 2
Note that we are using PW relations, because they are independent of
uncertainties affecting reddening corrections and also because they mimic
a Period-Luminosity-Color relation \citep{madore82,bono99c}.
We applied a 3$\sigma$ clipping to remove outliers and we found that the standard
deviations range from 0.08 mag for CCs to 0.10 mag for TIICs and to 0.13 mag
for RRLs. This result suggests a similar width in temperature of the instability
strip when moving from RRLs to CCs. 
       
We also note that the distribution of TIICs inside the      
instability strip is far from being homogeneous. Indeed, the current      
observations display two well defined clumps: a fainter one      
located at $\log L/L_\odot\approx$2.0 and $\log T_{eff}\approx$3.80      
and a brighter one located at $\log L/L_\odot\approx$2.6-2.8 and      
$\log T_{eff}\approx$3.74. The iso-periodic lines (black dashed lines)      
indicate that the former group is mainly associated with      
BLHs variables, given that their periods range from 1 to 5 days,      
while the latter one is associated to WVs variables with periods roughly      
ranging from 5 to 20 days. The predicted periods were estimated      
using the fundamental pulsation relation provided by \cite{marconi15}.      
%
% Ref. 5
This relation depends on stellar luminosity, effective temperature, 
stellar mass and chemical composition. For the first two parameters we are 
using individual values (HRD), while for the last two we are using plausible 
mean values. 
The correlation with the period distributions plotted in      
Fig.~\ref{fig_period} is quite obvious.    
     
%^^^^^^^^^^^^^^^^^^^^^^^^^^^^^^^^^^^^^^^^^^^^^^^^^^^^^^^^^^^^^^^^^^^^^^^^^^^^^^^     
% Ref. 3b
\subsection{Evolutionary properties of HB models}\label{sec:3.2}     
%^^^^^^^^^^^^^^^^^^^^^^^^^^^^^^^^^^^^^^^^^^^^^^^^^^^^^^^^^^^^^^^^^^^^^^^^^^^^^^^     

The evolutionary properties of low-mass core helium burning models have been      
discussed in several recent investigations      
\citep[][and references therein]{cassisi11}. Here we summarize the main      
features relevant to explain the evolutionary channels producing TIICs.         
     
The grey area displayed in the top panel of Fig.~\ref{theo_tracks} outlines    
the   region between the ZAHB (faint envelope) and central helium exhaustion      
(bright envelope) for a set of HB models with different masses and fixed    
chemical composition, Z=0.01 and He mass fraction Y=0.259.      
We have assumed an $\alpha$-enhanced chemical composition \citep{pietrinferni06b}    
and a progenitor mass according to a 13~Gyr isochrone (the mass at the main sequence    
turn off --MSTO-- is equal to 0.86 $M_\odot$).    
   
As well known, along the ZAHB the total mass of the models decreases when moving      
from the red HB (RHB) to the blue HB (BHB) and further on to the Extreme HB (EHB).      
The helium core mass is constant along the ZAHB and it is mainly fixed by the    
chemical composition  of the progenitor ($M_c^{He}$=0.4782 $M_\odot$) and is    
roughly independent of age for ages above a few Gyr. The mass of the envelope decreases      
from 0.4218$M_\odot$ for RHB models to a few thousandths of solar masses for EHB models.      
In an actual old stellar population with a fixed initial chemical composition, the mass    
lost along the RGB \citep[more efficient when approaching the tip of the RGB,]   
[]{origlia14} determines the final mass distribution along the ZAHB.

%^^^^^^^^^^^^^^^^^^^^^^^^^^^^^^^^^^^^^^^^^^^^^^^^^^^^^^^^^^^^^^^^^^^^^^^^^^^^^^^     
\begin{figure*}     
\centering     
\includegraphics[width=15cm, height=19truecm]{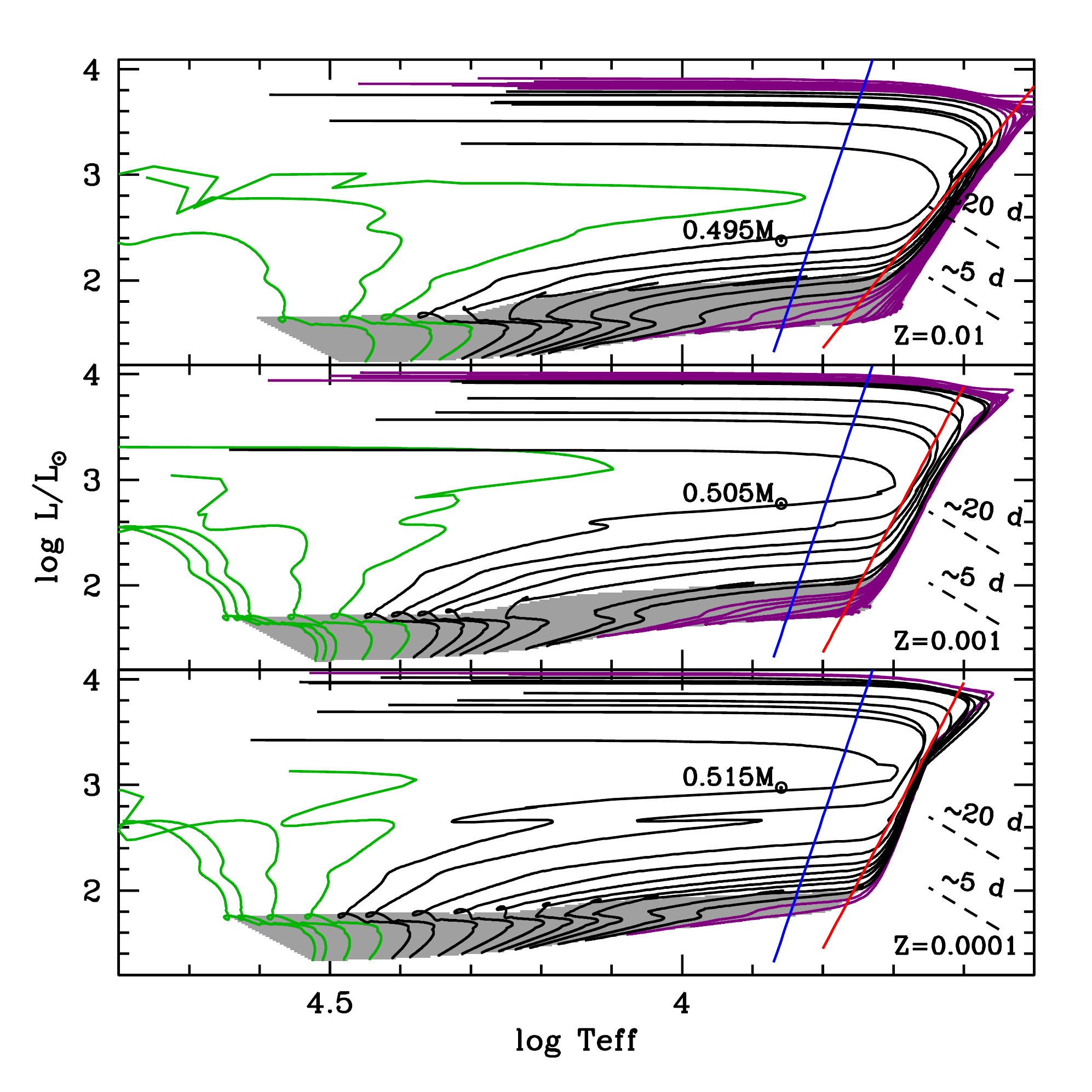}     
\caption{     
Top:  HRD of HB evolutionary models covering a broad range of stellar  masses    
(M/M$_\odot$=0.48--0.90)  and the same initial chemical composition    
(Z=0.01, Y=0.259). The grey area outlines the region between ZAHB      
(faint envelope) and central helium exhaustion (bright envelope).     
The green lines display HB models      
evolving as AGB-manqu\'e, black lines the post early-AGB models      
and purple lines the thermal pulsing AGB ones (see text for more details).       
The almost vertical blue and red solid lines mark the hot and cool edges       
of the Cepheid instability strip. The minimum stellar mass (solar      
units) crossing the instability strip is labelled in black.      
The black dashed lines display two iso-periodic lines for 5 and 20 days.     
%minima 0.480 0.90    
%    
Middle: Same as the top panel, but for stellar masses ranging from    
M/M$_\odot$=0.4912 to 0.80 and for a metal-intermediate chemical composition      
(Z=0.001, Y=0.246).     
Bottom: Same as the top panel, but for stellar masses ranging from    
M/M$_\odot$=0.5035 to 0.70 and for a metal-poor chemical composition      
(Z=0.0001, Y=0.245).    
}      
\label{theo_tracks}     
\end{figure*}     
%^^^^^^^^^^^^^^^^^^^^^^^^^^^^^^^^^^^^^^^^^^^^^^^^^^^^^^^^^^^^^^^^^^^^^^^^^^^^^^^     
     
The bright envelope of the grey area marks the central helium exhaustion,    
corresponding formally to the beginning of the AGB phase. The small ripples    
along the helium exhaustion sequence ($\log L/L_\odot\sim$1.8) display that    
the lower the total   mass of the HB model, the hotter the effective temperature    
at which the helium exhaustion takes place. The luminosity of the ripples      
ranges from $\log L/L_\odot\approx$2 in the warm region to      
$\log L/L_\odot\approx$1.6 in the hot region of the HB.      
At this point the He-burning moves smoothly to a shell around the carbon-oxygen core. The overlying H-shell    
extinguishes, due to the expansion of the structure before reigniting later with various efficiencies, depending on   
the mass thickness of the envelope around the original He core.   
   
Models with mass below 0.495$M_{\odot}$ (corresponding to an envelope mass lower than about 0.017$M_{\odot}$)   
never reach the AGB location, they do not cross the instability strip and move      
to their WD cooling sequence, as a carbon-oxygen      
(CO) WD \citep{castellani06,bono13,salaris13b}.     
These objects have been called AGB-Manqu\'e \citep{greggio90}, and are shown as green tracks in the top panel of      
Fig.~\ref{theo_tracks}).   
   
More massive models do cross the instability strip while moving towards their AGB tracks.    
Models with $0.495 \leq M/M_{\odot} < 0.55 $ do reach the AGB but move back towards    
the WD sequence (hence they cross the instability strip again but at higher    
luminosities) well before reaching the thermal pulse (TP) phase. They are named    
post early-AGB (PEAGB), and are plotted as black lines in the top panel    
of Fig.~\ref{theo_tracks}. These AGB models perform several gravo-nuclear    
loops in the HRD, either during the AGB phase and/or in their approach to    
the WD cooling sequence after leaving the AGB (during this post-AGB transition    
models cross again the instability strip). Some of them may take place inside the     
instability strip. The reader interested in a more detailed discussion      
concerning their impact on the pulsation properties is referred to      
\cite{bono97b,bono97d}. The evolutionary implications, and in particular the    
impact of the loops concerning the AGB lifetime have been recently discussed    
by \cite{constantino16}.   
   
Models with $M \geq 0.54 M_{\odot}$ (plotted as purple lines in Fig.    
\ref{theo_tracks}) do evolve along the AGB and experience the TPs. 
The number of TPs, and in turn,      
the duration of their AGB phase is once again dictated by the efficiency of     
the mass loss, and by their residual envelope mass \citep{weiss09,cristallo09}.     
Calculations of TP evolution are quite demanding from the computational point    
of view, hence  we decided to use the fast and simplified synthetic AGB    
technique originally developed by \cite{iben78} and more recently by    
\cite{wagenhuber98}, to compute the approach of these AGB models to the    
WD cooling sequence.    
In particular, the synthetic AGB modelling started for thermal-pulsing-AGB   
(TPAGB) models just before the occurrence of the first TP, while    
for PEAGBs it was initiated at the brightest and reddest point    
along the first crossing of the HRD, towards the AGB.   
       
The middle and the bottom panels of Fig.~\ref{theo_tracks} show the same      
predictions, but for two more metal-poor chemical compositions.      
The values of the stellar masses plotted along the ZAHBs show the      
impact of the chemical composition.   
The mass range of the tracks which cross the instability strip and produce TIICs steadily      
decrease from 0.495--0.90 $M/M_\odot$ for Z=0.01, to 0.505--0.80 $M/M_\odot$      
for Z=0.001 and to 0.515--0.70 $M/M_\odot$ for Z=0.0001. It is worth mentioning    
that the range in luminosity covered by the different sets of models is    
relatively similar. The mild change in stellar mass and the similarity in    
luminosities suggests a marginal dependence of the mass-luminosity (ML)    
relation of TIICs on the chemical composition.

%^^^^^^^^^^^^^^^^^^^^^^^^^^^^^^^^^^^^^^^^^^^^^^^^^^^^^^^^^^^^^^^^^^^^^^^^^^^^^^^     
\subsection{Pulsation and evolutionary implications for Type II Cepheids}\label{sec:1}     
%^^^^^^^^^^^^^^^^^^^^^^^^^^^^^^^^^^^^^^^^^^^^^^^^^^^^^^^^^^^^^^^^^^^^^^^^^^^^^^^     
     
The marginal dependence of the ML relation for TIICs on chemical composition      
does not imply similar period distributions in different stellar systems.   
To investigate in more detail this issue we plotted in Fig.~\ref{theo_tracks}   
the instability strips predicted by \citet{marconi15} for the various metal   
abundances.       
Data plotted in this figure show that the instability strip covers      
a broader range in effective temperatures when moving from metal-poor to      
metal-rich models. Moreover, HB evolutionary models become, as  expected,    
systematically redder when increasing the initial metallicity.    
These two circumstantial evidence indicates that both      
the period distribution and the evolutionary time spent inside the instability      
strip do depend on the metal content.        
     
A glance at the predictions plotted in this      
figure show that PEAGB models during either the first and partially    
during the second crossing of the instability strip can explain the    
typical period range of both BLHs and WVs. 
The increase of the pulsation period when moving from BLHs to WVs     
stems from the increase in luminosity and the decrease in stellar    
mass (see Fig.~\ref{theo_tracks}).     
% Ref. 4  
The evolutionary times spent inside the instability strip for the 
two crossings and for selected stellar masses are listed in 
Table~\ref{tab:tempievolutivi}. Note that the stellar masses, at fixed
chemical composition, were selected to be representative of the stellar
structures crossing the instability strip.
     
%%%%%%%%%%%%%%%%%%%%%%%%%%%%%%%%%%%%%%%%%%%%%%%%%%%%%%%%%%%%%%%%%%%%%%%%5  
% Table 2 
%%%%%%%%%%%%%%%%%%%%%%%%%%%%%%%%%%%%%%%%%%%%%%%%%%%%%%%%%%%%%%%%%%%%%%%%5  
\begin{table}     
\caption{Predicted population ratios based on synthetic AGB models 
by assuming an uncertainty of $\pm$50 K on the boundaries of the 
instability strip (see text for details).}     
\label{tab:popratio}     
\centering     
\begin{tabular}{lrccc}     
\hline     
Z & Ntot & NBLH/Ntot   &  NWV/Ntot  &    NRVT/Ntot \\     
\hline     
\multicolumn{5}{c}{---Uniform mass distribution---}\\     
0.0001 & 262$\pm$43 & 0.46$\pm$0.03 & 0.15$\pm$0.03 & 0.39$\pm$0.01 \\  
0.001  & 185$\pm$22 & 0.57$\pm$0.06 & 0.12$\pm$0.01 & 0.31$\pm$0.02 \\  
0.01   & 160$\pm$21 & 0.51$\pm$0.02 & 0.19$\pm$0.03 & 0.30$\pm$0.01 \\  
\multicolumn{5}{c}{---Bi-modal mass distribution---}\\     
0.0001 & 115$\pm$7   & 0.42$\pm$0.05  & 0.27$\pm$0.05  & 0.31$\pm$0.02  \\  
0.001  &  98$\pm$7   & 0.41$\pm$0.03  & 0.20$\pm$0.02  & 0.39$\pm$0.02  \\  
0.01   & 105$\pm$28  & 0.43$\pm$0.04  & 0.23$\pm$0.05  & 0.34$\pm$0.01  \\ 
\hline     
\end{tabular}     
%} 
\end{table}

To further define the theoretical framework for RVTs stars    
\citep{wallerstein2002,soszynski2011}, we suggest that they are    
the progeny of both PEAGB and TPAGB. Reasons supporting this hypothesis    
are the following:\par       
     
a) Period range -- The theoretical periods for these models are systematically      
longer than WVs and more typical of RVTs stars. The predicted mass for these      
structures is uncertain, because it depends on the efficiency of mass      
loss during the TP phase. The theoretical framework is further      
complicated by the fact that the number of TPs also depends on      
the initial mass of the progenitor and on its initial chemical composition.      
This means that it cannot be {\em a priori} excluded a contribution from      
intermediate-mass stars. However, the lack of RVTs in nearby dwarf      
spheroidal galaxies hosting a sizable fraction of intermediate-mass stars with      
ages ranging from 1~Gyr to more than 6-8~Gyrs such as      
(Carina, Fornax, Sextans; \cite{aparicio04,beaton2018}) is suggesting that this      
channel might not be very efficient. However, RVTs have been identified in      
the MCs \citep{soszynski08c,ripepi2015}.         
     
b) Alternating cycle behaviour -- There is evidence of an interaction      
between the central star and the circumstellar envelope possibly causing the      
alternating cycle behaviour \citep{feast08,rabidoux10}. The final crossing      
of the instability strip before approaching the WD cooling sequence either    
for PEAGB or for TPAGB models appears a very plausible explanation.    
     
The above circumstantial evidence suggests that the variable stars    
currently classified as TIICs have a range of evolutionary properties. The BLHs      
and the WVs appear to be either post-ZAHB (AGB, double shell burning) or      
post-AGB (shell hydrogen burning) stars, while RVTs are mainly       
post-AGB objects.  
     
Pulsation and evolutionary results plotted in Fig.~\ref{theo_tracks}      
display a few interesting features worth being discussed.      
     
We start with the ML relation, noticing that current models show       
that the mass of TIICs during the first crossing is steadily      
decreasing when we move from fainter to brighter structures.      
The opposite happens during the second crossing, i.e. the brighter the      
stellar structure, the larger the stellar mass. However, the difference in      
stellar mass along first and second crossing is smaller than 10\%. This is the    
reason why the pulsation period is steadily increasing over the entire    
luminosity range. This increase is driven by the increase in luminosity    
\citep[see the luminosity coefficients in the fundamental pulsation    
relation provided by][]{marconi15}.       
     
It is worth mentioning that the ML relation for TIICs is significantly      
different than for both CCs and RRLs. An increase in mass      
causes, for the former variables, an increase in the mean luminosity,      
and in turn, in the pulsation period. In the latter group the difference      
in mass is quite modest inside the instability strip, and the increase      
in pulsation period is mainly driven by a decrease in effective temperature.      
   
A second point to discuss concerns the pulsational period changes.    
The evolutionary framework we are outlining has some relevant consequences    
concerning the period changes of the variables. The      
BLHs should mainly display positive period changes, due to their      
main redward evolution. The only exceptions are in the short period regime      
since they evolve along their blueward excursion, and therefore they might      
experience both negative and positive period changes. The same behaviour is    
also expected for WV variables due to their main redward and partial blueward    
evolution. The main difference being that the redward evolution is    
systematically slower than the blueward one. Indeed, the duration of    
the first crossings of the instability strip are on average on the order of    
a few Myrs, while the second crossings are from a few times to one order of    
magnitude faster, depending on the stellar mass \citep{salaris08,bono13}.     
Therefore, the positive period changes should be on average more typical than      
the negative ones. Moreover, the negative period changes should mainly affect    
the long-period tail of WVs. These qualitative explanations rely on the    
assumption that these models do not experience gravo-nuclear loops inside    
the instability strip, which could significantly affect the previous inferences   
\citep{bono97b,bono97d}, with period changes   showing alternating positive    
and negative values.        
     
Finally, the RVTs should be mainly dominated by negative period changes,    
as a consequence of their blueward evolution.   This is true only in case    
their evolution is not affected by the TP phase. In this    
case they might experience both positive and negative period changes.       
       
%______________________________________________________________________________     
\section{Synthetic HB models}\label{sec:synthetic}     
%______________________________________________________________________________     
     
In the theoretical framework outlined in Section~\ref{sec:evolution},    
the post-early AGB stars produce BLHs and short period WVs when they are in    
the double shell   evolution and longer period WVs just before they approach    
their WD cooling sequence. Finally, TPAGBs evolving towards their   
WD cooling sequence produce long-period  RVTs. For a more quantitative   
analysis  we have computed three sets of synthetic models to account for the   
evolutionary times spent inside the predicted instability strips.      
The synthetic HB models have been computed with the code fully described   
in \citet{dalessandro13a}, and adapted to the problem at hand.
     
We decided to follow a simple approach: For each of the three initial chemical compositions    
of Fig.~\ref{theo_tracks}, we computed synthetic HB models employing the same tracks as in the figure, 
assuming at first       
an uniform mass distribution from the extreme HB to the red HB (the values of $Y$, $Z$ 
and mass ranges are given in the figure caption). 
In brief, the synthetic HBs were calculated as follows.

For a given $(Z,Y)$ pair (the He abundance is kept constant at fixed $Z$) 
we first select randomly (uniform probability) a value of the HB mass $m_s$ in the appropriate mass range, and the 
corresponding HB track is determined by interpolation in mass among the 
available set of HB tracks at that metallicity.

As a second step, the position of this HB mass $m_s$ in the HRD is determined according to its location
along the track 
after a HB evolutionary time $t_{ev}$ has been randomly extracted. With the standard assumption 
that stars are fed to the ZAHB at a constant rate, $t_{ev}$ is calculated by selecting randomly
(uniform probability) 
an age ranging from zero to $t_{tot}$, where $t_{tot}$ 
denotes the time spent from the ZAHB to the end of the computed evolutionary sequence with the longest lifetime.
This implies that
for some value of $m_s$ the randomly selected $t_{ev}$ will be
longer than the lifetime covered by the corresponding track or, in other words, that this mass has already 
evolved beyond the last evolutionary point covered by the calculations. 
Finally, to properly sample fast evolutionary phases, we included in our simulations a large number of synthetic stars (N=50,000). As we have tested by altering the value of N, this number is high enough to ensure that the
derived number ratios of the variables are statistically robust.
  
For each simulation Fig.~\ref{sinteticgap} shows the HRD of the synthetic stars.    
The stellar structures located inside the instability strip      
and producing TIICs were marked with asterisks.

%^^^^^^^^^^^^^^^^^^^^^^^^^^^^^^^^^^^^^^^^^^^^^^^^^^^^^^^^^^^^^^^^^^^^^^^^^^^^^^^     
\begin{figure}[!htbp]     
\centering     
\includegraphics[width=8.5cm,  height=10cm]{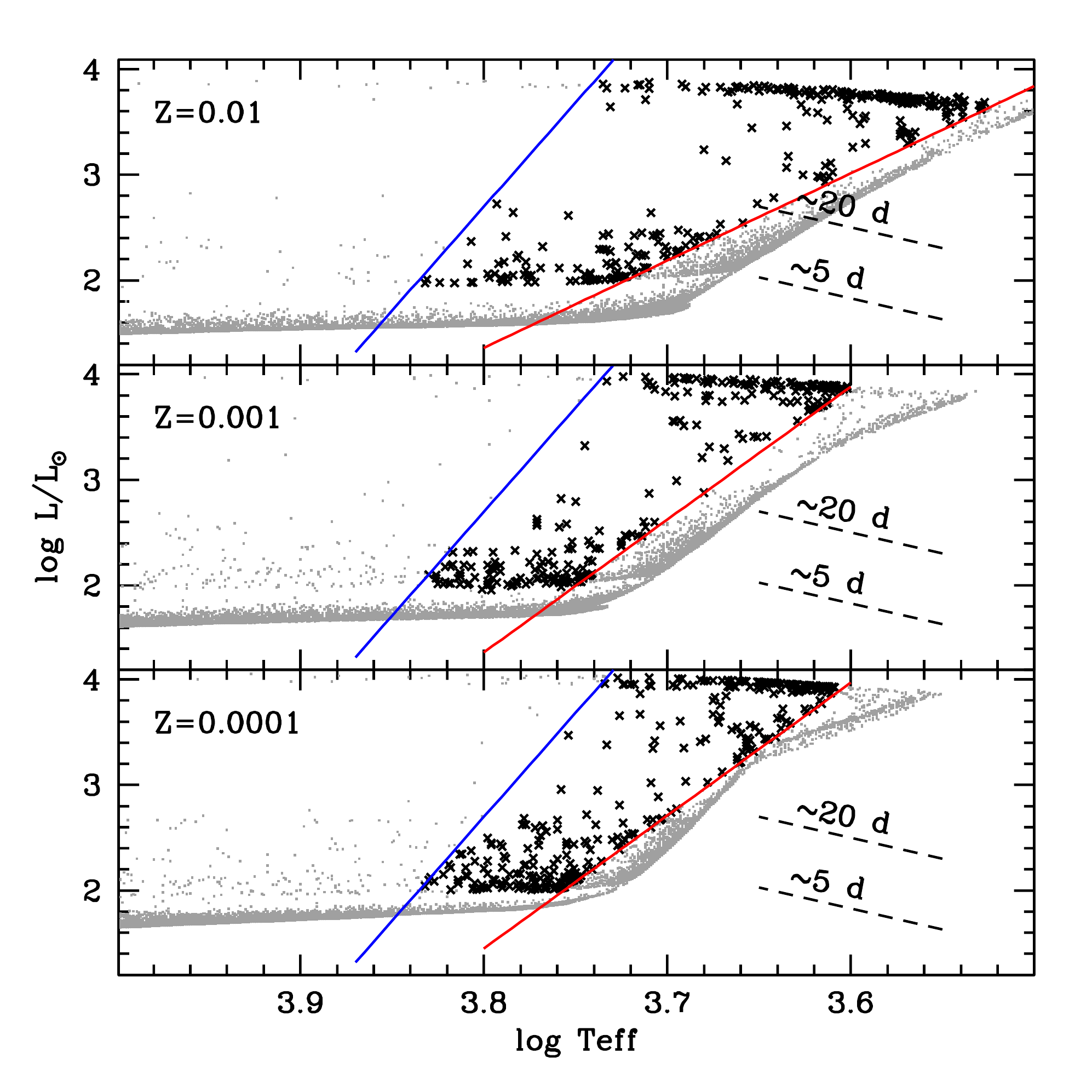}     
\caption{Same as Fig.~\ref{theo_tracks}, but for synthetic HB models.       
Objects inside the instability strip and      
producing TIICs are marked with asterisks. Those either falling outside      
the instability strip or producing RRLs are marked with grey dots.}     
\label{sinteticgap}     
\end{figure}     
%^^^^^^^^^^^^^^^^^^^^^^^^^^^^^^^^^^^^^^^^^^^^^^^^^^^^^^^^^^^^^^^^^^^^^^^^^^^^^^^     
   
%^^^^^^^^^^^^^^^^^^^^^^^^^^^^^^^^^^^^^^^^^^^^^^^^^^^^^^^^^^^^^^^^^^^^^^^^^^^^^^^     
\begin{figure}[!htbp]     
\centering     
\includegraphics[width=8.5cm, height=10cm]{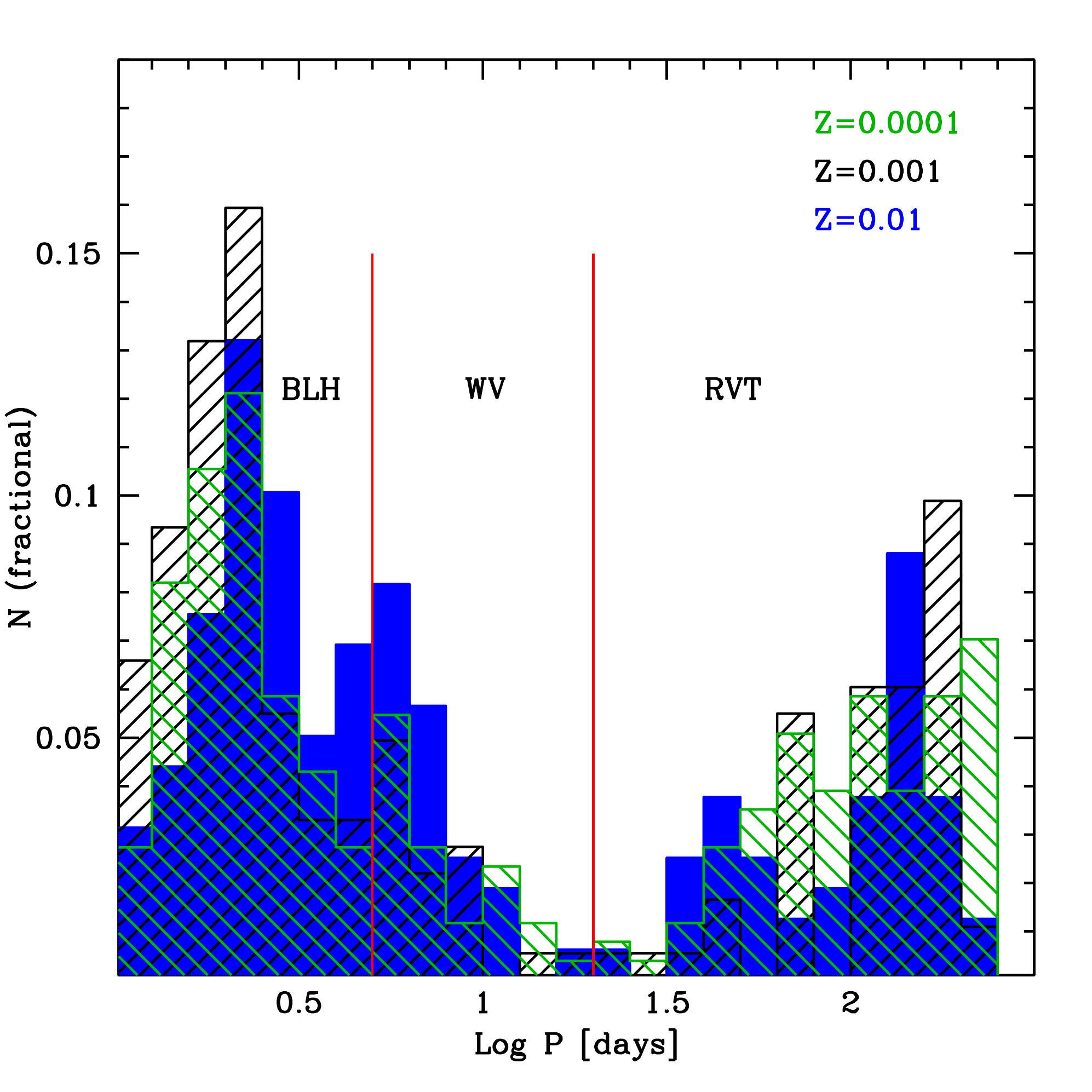}     
%\vspace{-2cm}     
\caption{Predicted period distributions for TIICs with different chemical      
compositions (the colour coding is labeled). Results are based on synthetic      
HB models (see Fig.~\ref{theo_tracks}) calculated by assuming an uniform mass distribution     
along the HB. The two vertical red lines display the boundaries at 5      
and 20 days, respectively. The different subgroups of TIICs are also labelled.}     
\label{per_uni}     
\end{figure}     
%^^^^^^^^^^^^^^^^^^^^^^^^^^^^^^^^^^^^^^   
   
%______________________________________________________________________________     
\subsection{Predicted period distributions}\label{sec:synthetic}     
%______________________________________________________________________________     
Figure 7 shows the predicted period distributions, based on the fundamental  
pulsation relation by \citet{marconi15}, for the three adopted  
chemical compositions.   
The TIIC pulsation properties marginally depend on the metal content, in   
fair agreement with observations.   
Metal-rich (blue hatched area) models show a more pronounced peak at   
the transition between BLHs and WVs compared to observations, while metal-intermediate (black hatched area)   
and metal-poor (green hatched area) models display a small difference in the   
short-period tail ($\log P\sim$0.15) for the BLHs and in the long-period tail   
for RVs ($\log P\sim$2.3). However, predictions based on an uniform mass   
distribution shows two discrepancies with observations.   
i) The period distributions for RVTs attain their maxima at periods   
longer than 100 days, while data plotted in Figures 2 and 3 display only   
a few objects in this range.   
ii) The empirical periods in the WV regime display a well defined peak      
at values ranging from 12 to 16 days. On the other hand, the corresponding theoretical      
predictions display a peak at periods around six      
days and a steady decrease in approaching the transition between   
WVs and RVTs.

%%%%%%%%%%%%%%%%%%%%%%%%%%%%%%%%%%%%%%%%%%%%%%%%%%%%%%%%%%%%%%%%%%%%%%%%5  
% Table 3 
%%%%%%%%%%%%%%%%%%%%%%%%%%%%%%%%%%%%%%%%%%%%%%%%%%%%%%%%%%%%%%%%%%%%%%%%5  
\begin{table}     
\scriptsize    
\caption{Evolutionary times spent inside the instability strip for      
different HB models constructed assuming, respectively,   
synthetic AGB modelling, full evolutionary calculations (Gravo--nuclear 
loops), and He-enhanced (Y=0.30) chemical composition.}       
\label{tab:tempievolutivi}     
\centering     
\begin{tabular}{llccc}     
\hline     
\hline     
Z       & Stellar Mass & Synthetic HB & Gravo--nuclear loops & He-enriched \\     
        & [M$_\odot$]  &     [Myr]    &        [Myr]         &    [Myr]    \\     
\hline     
0.0001  & 0.525 & 1.28   &  0.51 &  3.19 \\     
0.001   & 0.525 & 1.10   &  0.28 &  2.59\\     
0.01    & 0.505 & 4.05   &  3.2  &  5.66\\     
\hline     
\end{tabular}     
\end{table}     
     
To quantify the dependence of the various types of variables on the      
chemical composition, we estimated the population ratios, i.e. the   
number of BLHs, WVs and RVTs over the total number of TIICs. Values   
listed in Table~\ref{tab:popratio} show       
that the population ratios show marginal variations, when moving   
from the metal-rich to the more metal-poor regime.   
However, the population ratios for BLHs are 20\% larger than   
observed, while those for WVs and RVTs are almost three times smaller   
and two times larger than observed, respectively (see Fig.~\ref{fig_period}).  
Note that the standard deviations associated to the population ratios 
listed in Table~\ref{tab:popratio} take into account uncertainties 
(by $\pm$50 K) on the effective temperature of the predicted boundaries 
of the instability strip.  
     
%______________________________________________________________________________     
\subsection{Bimodal mass distribution along the HB}\label{sec:synthetic}     
%______________________________________________________________________________     
     
Theory and observations \citep{castellani2003} suggest that the mass    
distribution along the HB is far from being uniform.   
Moreover, GCs showing extended HBs also display well-defined   
gaps in the HB luminosity function \citep{castellani2007,torelli2019}.  
To investigate the impact of a non-uniform mass distribution on the predicted periods,    
we computed synthetic HB models for a metal-poor chemical  composition   
(Z=0.0001) and a bimodal Gaussian mass distribution centred on two   
different mean stellar masses    
(M=0.535, 0.57 $M_\odot$), with the same standard deviation    
($\sigma$=0.005 $M_\odot$).
These simulations are calculated in exactly the same way as for the 
case of a uniform mass distribution.  The only difference is that in 
the first step of the calculations, the masses $m_s$ of the synthetic stars 
are now randomly selected according to Gaussian distributions with 
prescribed mean values and $\sigma$ dispersions. 

We performed a number of numerical experiments    
by changing the two mean values to obtain population ratios and period    
distributions of variables similar to the observed ones. The period distribution based on    
metal-poor synthetic HB models is plotted as a green hatched histogram in    
Fig.~\ref{per_104}.      
The new period distribution globally agrees with observations, indeed, it      
naturally shows a double peak. The peak in the      
BLH regime agrees quite well with observations, and the whole period distribution      
covers a similar range in periods. The gap between BLH and WVs appears to      
be located around four--five days, instead of five days. The main difference      
is in the WV regime, where the predicted peak      
in the period distribution is located around six days instead of 12-16 days.      
The increase either by a factor of ten (black hatched histogram in    
Fig.~\ref{per_104}) or by a factor one hundred (blue hatched histogram in    
Fig.~\ref{per_104}) in metal content has a marginal impact on the position    
of the peak in the period distribution of WVs.  
Notice that the metal-intermediate and metal-rich synthetic HB models, to   
match observational constraints coming from different stellar populations,   
have been calculated with different choices of the mean stellar    
masses (see labeled values), but the same standard deviation ($\sigma$=0.005 $M_\odot$).       
The discrepancy concerning the predicted period distribution for RVTs appears to be   
partially mitigated, because they move towards shorter periods. However,   
predicted periods for RVTs are systematically longer than observed.     
  
Interestingly, the population ratios listed in   
Table~\ref{tab:popratio} show that bimodal mass distributions along   
the HB take account for the observed ratio for BLHs. The predicted population   
ratios for WVs and RVTs are still smaller and larger than observed, respectively, 
but the difference between theory and observations is within a factor of two.

%^^^^^^^^^^^^^^^^^^^^^^^^^^^^^^^^^^^^^^^^^^^^^^^^^^^^^^^^^^^^^^^^^^^^^^^^^^^^^^^     
\begin{figure}     
\centering     
\includegraphics[width=8.5cm, height=12cm]{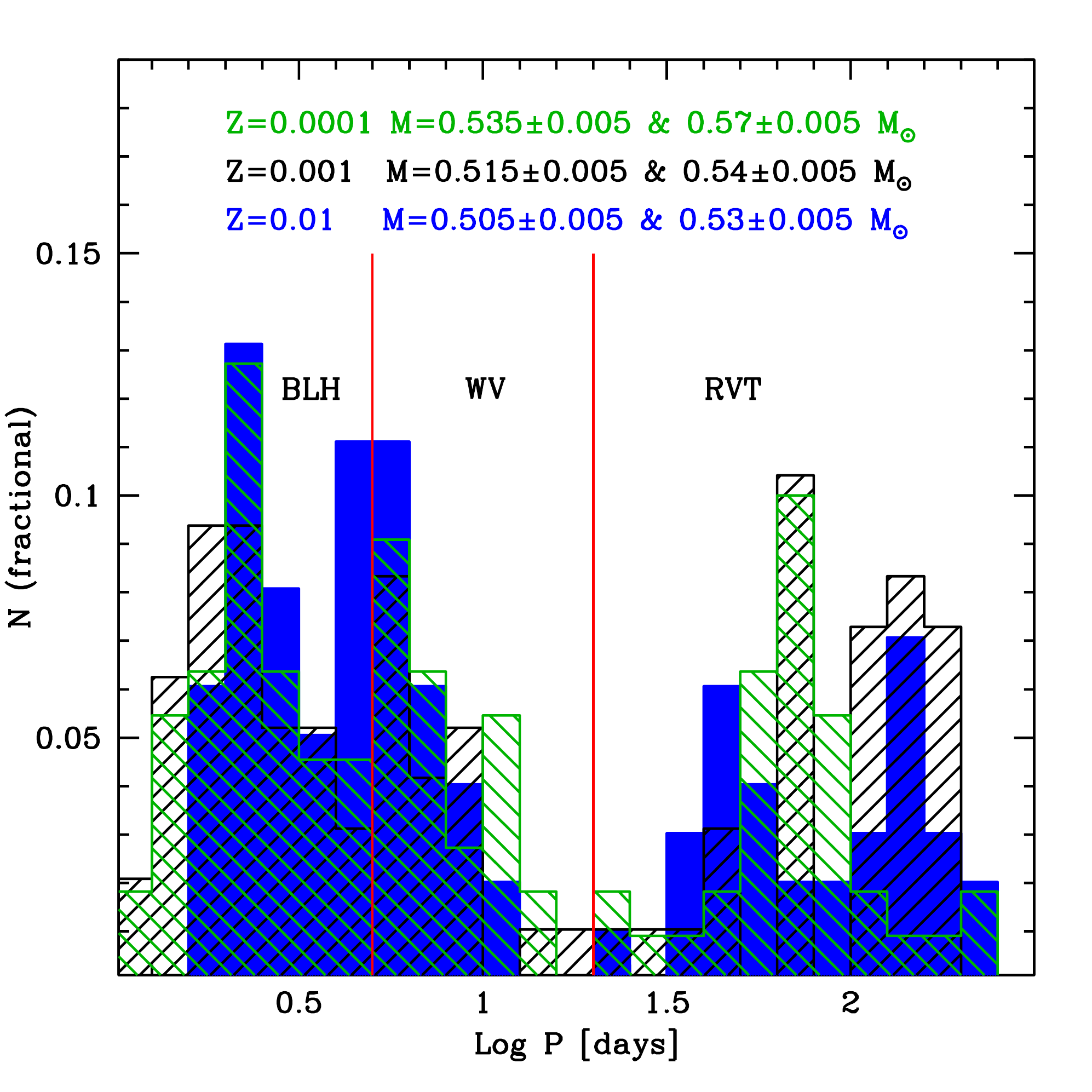}     
\caption{Same as Fig.~\ref{per_uni}, but for synthetic HB models which assume a   
bimodal HB mass distribution and the values of the adopted mean masses are 
labeled. The standard deviations for the mass distributions is the same and 
equal to 0.005~M$_\odot$. The two vertical blue      
lines display the boundaries at five and twenty days, respectively.     
}     
\label{per_104}     
\end{figure}     
%^^^^^^^^^^^^^^^^^^^^^^^^^^^^^^^^^^^^^^^^^^^^^^^^^^^^^^^^^^^^^^^^^^^^^^^^^^^^^^^     
     
%______________________________________________________________________________     
\section{Detailed evolutionary calculations for off-ZAHB evolution}\label{sec:synthetic}     
%______________________________________________________________________________     
     
A detailed comparison between theory and observations would require      
more detailed synthetic modelling that accounts for a broad range of      
chemical compositions, progenitor masses and actual HB mass distributions.    
However, this approach is beyond the aim of the current investigation.    
We are interested here in providing a global theoretical framework for    
TIICs, and in particular, to trace the impact that different parameters    
play in their evolutionary and pulsation properties.

%^^^^^^^^^^^^^^^^^^^^^^^^^^^^^^^^^^^^^^^^^^^^^^^^^^^^^^^^^^^^^^^^^^^^^^^^^^^^^^^     
\begin{figure*}     
\centering     
\includegraphics[width=8cm, height=9cm]{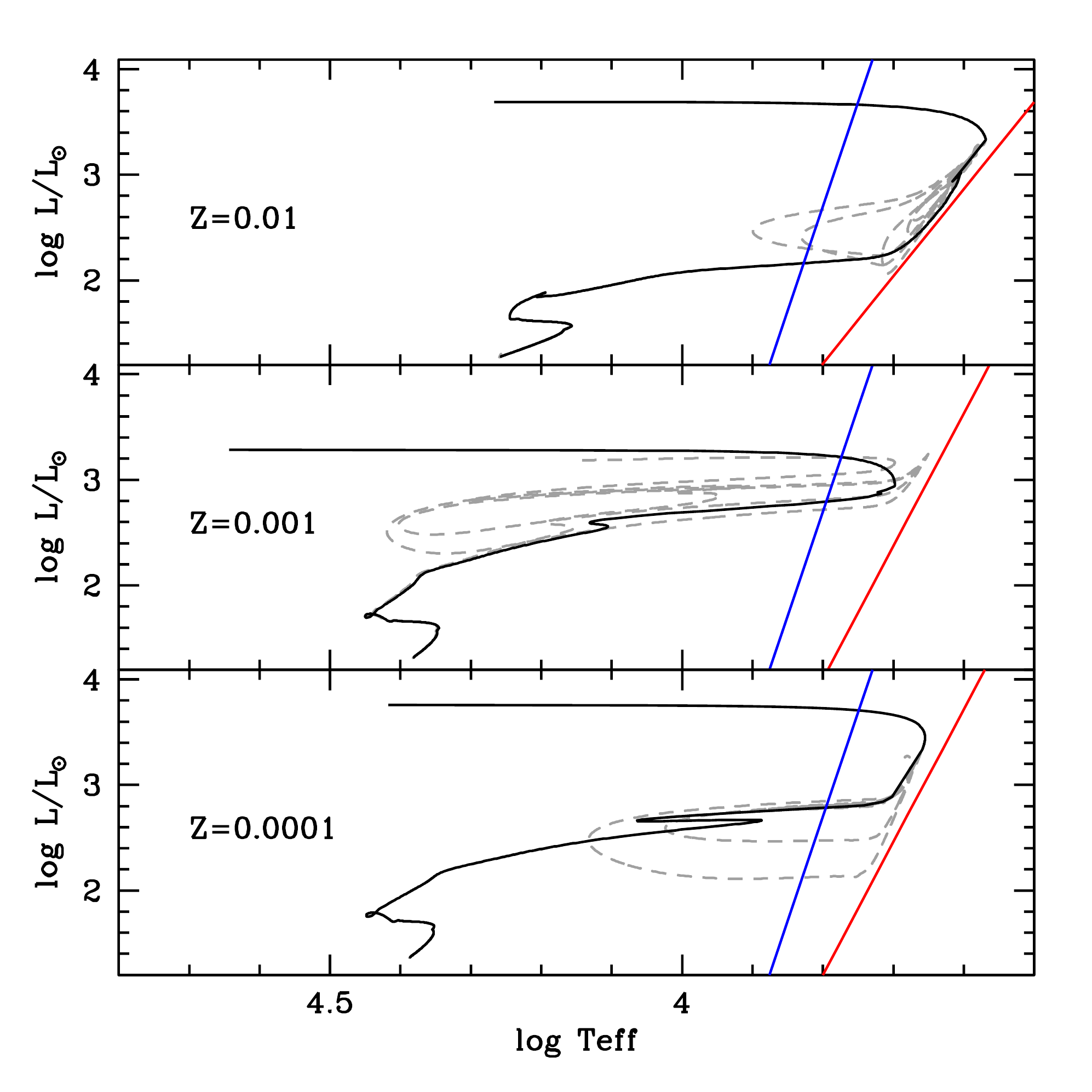}     
\includegraphics[width=8cm, height=9cm]{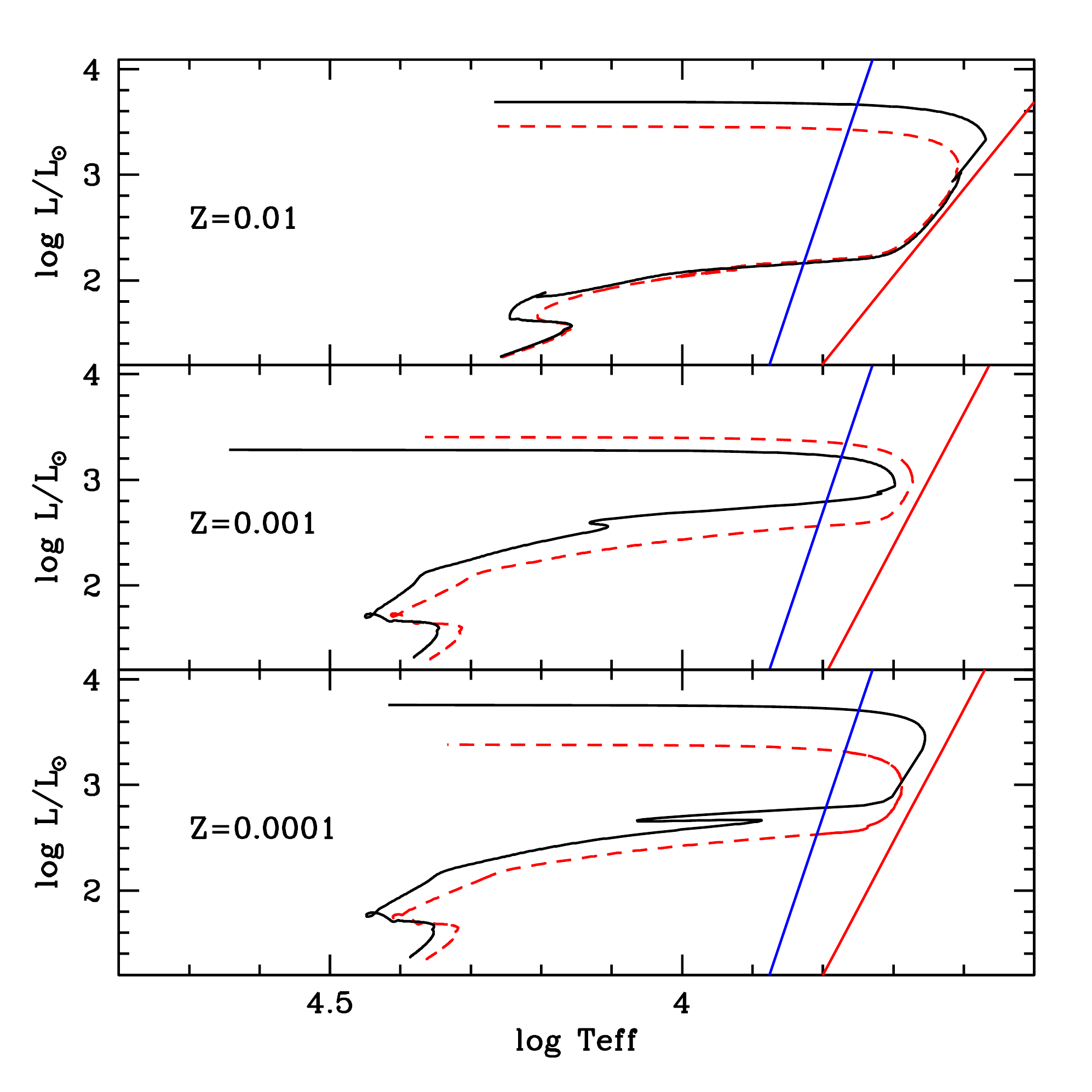}     
\caption{Left: HRD of HB models crossing the      
instability strip, calculated by assuming an analytical extension      
for the AGB evolutionary phases (black solid lines) and HB models      
with the same stellar mass with a full evolutionary computation (dashed black line).    
From top to bottom the theoretical predictions refer to metal-rich (M=0.505M$_\odot$),      
metal-intermediate (M=0.505M$_\odot$) and metal-poor (M=0.520M$_\odot$) models.   
The almost vertical blue and red lines display the edges of the      
instability strip.      
Right: Same as the left panel, but the comparison is between standard He (black solid line)      
and He-enhanced (Y=0.30) models (dashed red lines). See text for details.       
}     
\label{cozze}     
\end{figure*}     
%^^^^^^^^^^^^^^^^^^^^^^^^^^^^^^^^^^^^^^^^^^^^^^^^^^^^^^^^^^^^^^^^^^^^^^^^^^^^^^^     
     
As a preliminary step in this direction we have computed       
in detail the off-ZAHB evolution of some PEAGB models.      
The left panel of Fig.~\ref{cozze} shows the difference in the off-ZAHB      
evolution between HB models constructed by assuming the analytical AGB evolution      
(solid black line) and HB models with the same stellar mass (dashed black line)      
computed with full evolutionary calculations.    
The latter one performs, as expected, a series of gravo-nuclear loops inside      
the instability strip \citep{bono97e}. The number can range from a few up to  
several tens, and depends on the value of the stellar mass and on the chemical composition.      
Notice that the fully evolutionary models were integrated in time until the mass of    
the envelope surrounding the He core was smaller than 0.01 M$_\odot$.

The change in the evolutionary path has a twofold impact on the predicted    
period distribution.      
a) The loops took place in a luminosity regime ($\log L$/$L_{\odot}$=2.4-3.0)      
typical of WVs. This means that they mainly affect the shape and      
position of the peak in the period distribution located around 10-16 days.       
b) The evolutionary time spent in this luminosity regime is, on average,      
several Myrs longer than the evolutionary time based on analytical extensions      
of AGB evolution (see Figures 6 and 7 in \citealt{bono97e} and Figures 7-10      
in \citealt{constantino16}). Note that a significant portion of the red tips      
of the loops both for the metal-poor and the metal-intermediate chemical      
composition fall inside the instability strip.       
     
The quoted evidence indicates that direct calculations of low-mass AGB stars      
appear very promising to constrain the period distribution of WVs. Note      
that gravo-nuclear loops do not affect evolutionary properties of BLHs.      
The current evolutionary prescriptions indicate that they are not produced      
by HB stellar structures with more massive envelopes.       
     
%______________________________________________________________________________     
\subsection{Dependence of period distribution on primordial helium content}\label{sec:synthetic}     
%______________________________________________________________________________     
     
Finally, we have also investigated the dependence of the TIIC period distribution      
on the initial He content. The right panel of Fig.~\ref{cozze} shows the      
same HB models (black solid lines, AGB analytical extension)      
plotted in the left panel, but compared with Y=0.30, He enhanced tracks   
(and analytical extension of AGB evolution, dashed red line).   
The enhancement of helium causes an increase of the time spent inside the    
instability strip ranging from 30\% for the metal-rich models, to more than a factor    
of two for more metal-poor models (see values listed in Table~\ref{tab:tempievolutivi}).    
However, synthetic HB models are required to quantify   
the impact of the helium content on the period distributions.      
Note that detailed calculations for RRLs in \wcen show that an increase      
in helium mainly causes a systematic shift of the period distribution toward      
longer periods \citep{marconi15}.       
     
%______________________________________________________________________________     
\section{Summary and final remarks}\label{sec:summary}     
%______________________________________________________________________________     
     
We have reviewed the evolution and the pulsation properties of TIICs, and   
propose  a new evolutionary characterization of these pulsators.   
They are mainly old, low-mass stars    
either during the AGB (double shell burning) or post-AGB (hydrogen shell    
burning) evolutionary phases. In particular, BLHs are envisaged to be mainly   
PEAGB stars that  are still approaching their AGB track, while WVs are a mix   
of PEAGB and  post-AGB stars along their second crossing of the instability   
strip (moving from the cool to the hot side of the HRD).   
This indicates that BLHs and WVs share the same evolutionary channel,   
therefore, they should be considered a single group of variable stars.    
The RVTs are predicted to be a mix between post-AGB stars along their second   
crossing  (short-period tail) and TPAGB stars (long-period tail)   
moving  towards the WD cooling sequence.            
From the evolutionary point of view, TIICs are much more complex than RRLs    
and CCs, which are in their core He burning plus shell  H~burning phase.   
Moreover, it is not clear yet the role that binarity plays in shaping   
their properties.   
   
Moreover, the current theoretical framework suggests that blue (low-mass)   
HB stars in their first crossing of the instability strip produce both   
BLHs and WVs. This means that these variables are not associated with TPAGB stars   
crossing the instability strip during the so-called \lq{blue nose\rq}    
(or \lq{Gingold's nose\rq}).   
Moreover, it soundly supports the very first dynamical mass measured by   
\citet{pilecki2018}  by using a double-lined binary system including   
a TIICs (M=0.64$\pm$0.02 M$_\odot$).    
  
Futhermore, we addressed the following key points.      
     
$\#$ {\em Periods}-- The observed period distributions of TIICs in different stellar    
systems (Galactic bulge, Galactic globular clusters, Magellanic Clouds) display      
a broad similarity. There are small differences concerning the positions      
of the gaps in the distributions, but it is not clear whether they are a    
consequence of the environment (cluster versus field) or possible    
observational biases.     
     
$\#$  {\em Amplitudes}-- The observed luminosity amplitudes (Bailey diagram) of TIICs show a well      
defined double-peaked distribution.  The peaks cover a broad range in period      
and the amplitudes, at fixed pulsation period, display a large spread.      
     
$\#$  {\em Topology of the instability strip}-- Theory and observations       
show that TIICs mainly pulsate in the fundamental mode, because they reach       
luminosities brighter than the {\rm transition point}. The possible occurrence   
of fainter TIICs evolving along the blueward excursion showed by some HB models,   
and in turn the presence of first overtone pulsators   
\citep{soszynski2019b}, cannot be excluded.          
     
$\#$ {\em Pulsation characterization}-- On the basis of pulsational properties    
(luminosity amplitudes, periods), TIICs have been usually    
partitioned into three different sub-groups, namely BLHs, WVs and RVTs.    
There is no consensus on the criteria adopted to separate the    
different sub-groups. The lack of TIICs in nearby dwarf galaxies and    
the small number of TIICs in globulars hamper more quantitative    
constraints.      
     
$\#$ {\em Period changes}-- Our evolutionary scenario predicts       
both positive and negative period changes. Theoretical evolutionary models      
show the possible occurrence of multiple gravo-nuclear loops at the onset    
of He shell burning. These loops could also take place inside the instability    
strip, and therefore, they can cause faster period changes.        
     
$\#$ {\em Metallicity distribution}-- Observations show that the metallicity      
distribution of TIICs is, as expected, similar to field and cluster RRLs.    
However, the spectroscopic sample is still too limited ($\approx$140)    
to draw firm conclusions.        
     
$\#$ {\em Population ratios and period distributions}-- The observed population ratios    
of BLHs, WVs and RVTs are quite similar when moving      
from Galactic stellar systems (globulars, field) to the Magellanic Clouds, suggesting      
a common evolutionary channel. Theoretical predictions based on synthetic HB models   
With bi-modal mass distribution agree well with   
observed population ratios for BLHs. However, the predicted ratios for   
WVs and RVTs are almost a factor of two smaller and larger than observed, respectively. Moreover,   
the predicted  period distributions of WVs peak at periods systematically   
shorter than observed  (6 versus 12-16 days). The predicted period distributions   
for RVTs display a long period tail not supported by observations.   
We performed preliminary calculations that model in detail the    
gravo-nuclear loops during the early AGB phase, and calculations for He-enhanced compositions.        
Gravo-nuclear loops cause an increase of the time spent inside the instability strip    
(from a few to several Myrs longer than calculations based on synthetic AGB modelling).   
Similar increase in evolutionary times applies to He-enhanced models, but the increase    
ranges from 30\% (metal-rich stellar structures) to roughly a factor of two (more    
metal-poor stellar structures). Gravo-nuclear loops might explain the      
difference in the population ratio and in the period distribution of WVs. The current evolutionary      
and pulsation prescriptions indicate that an increase of the helium content mainly causes a shift of      
the period distribution toward longer periods. However, detailed synthetic HB models taking into    
account a variety of evolutionary scenarios are required to constrain observed properties    
of TIICs.   
     
$\#$ {\em Standard candles}-- The TIICs are ``lato sensu'' solid distance indicators,      
because they obey a well defined optical and NIR PL relations.   
Theory and observations       
suggest also that they are minimally/marginally affected by the initial metallicity.    
Furthermore, they also obey optical and NIR PW relations. The key advantage     
of these diagnostics is to be independent on reddening uncertainties, but they rely on      
the assumption that the reddening law is universal.        
This suggest that TIICs can be considered also ``stricto sensu''      
ideal distance indicators, able to provide very accurate relative distances,   
independent of uncertainties in the zero-point of the      
adopted PL relations. The vanishing dependence of the slope of optical      
and NIR PW relations further supports the use of these variables in stellar      
systems affected by differential reddening \citep{braga15}.     
Moreover, the uncertainties affecting the zero-points of these diagnostics and, in turn,      
the absolute distances based on TIICs are going to vanish in a few years, thanks      
to the very accurate trigonometric parallaxes that the Gaia mission is going to provide.       
     
The observational outlook for TIICs appear even more promising when we consider   
that JWST is going to provide a complete census of TIICs across Local Group      
and Local Volume galaxies. The haloes of giant galaxies are, indeed, marginally affected      
by crowding problems. Moreover, ELTs will provide the unique opportunity to trace old      
stellar populations \citep{fiorentino17} even in the bulges and in the innermost regions   
of nearby Universe   thanks to their superb spatial resolution.        
     
\section*{Acknowledgements}     
     
It is a real pleasure to thank G. Wallerstein for many useful discussions   
and suggestions concerning type II Cepheids. It is also a pleasure to thank   
the anonymous referee for her/his valuable suggestions that helped us to   
improve the content and the readability of the paper.   
GB thanks A Severo Ochoa research grant at the the Instituto de Astrofisica      
de Canarias, where part of this manuscript was written.       
GF has been supported by the Futuro in Ricerca 2013 (grant     
RBFR13J716).     
This research has been supported by the Spanish Ministry of Economy and      
Competitiveness (MINECO) under the grant AYA2014-56795-P.     
VFB and MF acknowledge the financial support of the Istituto Nazionale di   
Astrofisica (INAF), Osservatorio Astronomico di Roma, and   
Agenzia Spaziale Italiana (ASI) under contract to INAF:   
ASI 2014-049- R.0 dedicated to SSDC.     
     
%%%%%%%%%%%%%%%%%%%%%%%%%%%%%%%%%%%%%%%%%%%%%%%%%%     
     
     
     
\bibliographystyle{aa}

\begin{thebibliography}{140}
\expandafter\ifx\csname natexlab\endcsname\relax\def\natexlab#1{#1}\fi

\bibitem[{{Aparicio} \& {Gallart}(2004)}]{aparicio04}
{Aparicio}, A. \& {Gallart}, C. 2004, \aj, 128, 1465

\bibitem[{{Baade}(1958)}]{baade58a}
{Baade}, W. 1958, \aj, 63, 207

\bibitem[{{Baker} \& {Gough}(1979)}]{baker1979}
{Baker}, N.~H. \& {Gough}, D.~O. 1979, \apj, 234, 232

\bibitem[{{Beaton} {et~al.}(2018){Beaton}, {Bono}, {Braga}, {Dall'Ora},
  {Fiorentino}, {Jang}, {Mart{\'\i}nez-V{\'a}zquez}, {Matsunaga}, {Monelli},
  {Neeley}, \& {Salaris}}]{beaton2018}
{Beaton}, R.~L., {Bono}, G., {Braga}, V.~F., {et~al.} 2018, \ssr, 214, 113

\bibitem[{{Becker} {et~al.}(1977){Becker}, {Iben}, \& {Tuggle}}]{becker77}
{Becker}, S.~A., {Iben}, Jr., I., \& {Tuggle}, R.~S. 1977, \apj, 218, 633

\bibitem[{{Behr}(2003)}]{behr03}
{Behr}, B.~B. 2003, \apjs, 149, 67

\bibitem[{{Bhardwaj} {et~al.}(2017){Bhardwaj}, {Macri}, {Rejkuba}, {Kanbur},
  {Ngeow}, \& {Singh}}]{bhardwaj17b}
{Bhardwaj}, A., {Macri}, L.~M., {Rejkuba}, M., {et~al.} 2017, \aj, 153, 154

\bibitem[{{Bono} {et~al.}(1997{\natexlab{a}}){Bono}, {Caputo}, {Cassisi},
  {Incerpi}, \& {Marconi}}]{bono97b}
{Bono}, G., {Caputo}, F., {Cassisi}, S., {Incerpi}, R., \& {Marconi}, M.
  1997{\natexlab{a}}, \apj, 483, 811

\bibitem[{{Bono} {et~al.}(2000){Bono}, {Caputo}, {Cassisi}, {Marconi},
  {Piersanti}, \& {Tornamb{\`e}}}]{bono00b}
{Bono}, G., {Caputo}, F., {Cassisi}, S., {et~al.} 2000, \apj, 543, 955

\bibitem[{{Bono} {et~al.}(1997{\natexlab{b}}){Bono}, {Caputo}, {Castellani}, \&
  {Marconi}}]{bono97d}
{Bono}, G., {Caputo}, F., {Castellani}, V., \& {Marconi}, M.
  1997{\natexlab{b}}, \aaps, 121, 327

\bibitem[{{Bono} {et~al.}(1999){Bono}, {Caputo}, {Castellani}, \&
  {Marconi}}]{bono99b}
{Bono}, G., {Caputo}, F., {Castellani}, V., \& {Marconi}, M. 1999, \apj, 512,
  711

\bibitem[{{Bono} {et~al.}(2008){Bono}, {Caputo}, {Fiorentino}, {Marconi}, \&
  {Musella}}]{bono08}
{Bono}, G., {Caputo}, F., {Fiorentino}, G., {Marconi}, M., \& {Musella}, I.
  2008, \apj, 684, 102

\bibitem[{{Bono} {et~al.}(1997{\natexlab{c}}){Bono}, {Caputo}, \&
  {Santolamazza}}]{bono97e}
{Bono}, G., {Caputo}, F., \& {Santolamazza}, P. 1997{\natexlab{c}}, \aap, 317,
  171

\bibitem[{{Bono} {et~al.}(2016){Bono}, {d}, {Marconi}, {Braga}, {Fiorentino},
  {Stetson}, {Buonanno}, {Castellani}, {Dall'Ora}, {Fabrizio}, {Ferraro},
  {Giuffrida}, {Iannicola}, {Marengo}, {Magurno}, {Martinez-Vazquez},
  {Matsunaga}, {Monelli}, {Neeley}, {Rastello}, {Salaris}, {Short}, \&
  {Stellingwerf}}]{bono2016b}
{Bono}, G., {d}, A., {Marconi}, M., {et~al.} 2016, Commmunications of the
  Konkoly Observatory Hungary, 105, 149

\bibitem[{{Bono} \& {Marconi}(1999)}]{bono99c}
{Bono}, G. \& {Marconi}, M. 1999, in IAU Symposium, Vol. 190, New Views of the
  Magellanic Clouds, ed. Y.~H. {Chu}, N.~{Suntzeff}, J.~{Hesser}, \&
  D.~{Bohlender}, 527

\bibitem[{{Bono} {et~al.}(2013){Bono}, {Salaris}, \& {Gilmozzi}}]{bono13}
{Bono}, G., {Salaris}, M., \& {Gilmozzi}, R. 2013, \aap, 549, A102

\bibitem[{{Bono} \& {Stellingwerf}(1994)}]{bono94b}
{Bono}, G. \& {Stellingwerf}, R.~F. 1994, \apjs, 93, 233

\bibitem[{{Braga} {et~al.}(2019){Braga}, {Contreras Ramos}, {Minniti},
  {Ferreira Lopes}, {Catelan}, {Minniti}, {Nikzat}, \& {Zoccali}}]{braga2019b}
{Braga}, V.~F., {Contreras Ramos}, R., {Minniti}, D., {et~al.} 2019, \aap, 625,
  A151

\bibitem[{{Braga} {et~al.}(2015){Braga}, {Dall'Ora}, {Bono}, {Stetson},
  {Ferraro}, {Iannicola}, {Marengo}, {Neeley}, {Persson}, {Buonanno},
  {Coppola}, {Freedman}, {Madore}, {Marconi}, {Matsunaga}, {Monson}, {Rich},
  {Scowcroft}, \& {Seibert}}]{braga15}
{Braga}, V.~F., {Dall'Ora}, M., {Bono}, G., {et~al.} 2015, \apj, 799, 165

\bibitem[{{Braga} {et~al.}(2016){Braga}, {Stetson}, {Bono}, {Dall'Ora},
  {Ferraro}, {Fiorentino}, {Freyhammer}, {Iannicola}, {Marengo}, {Neeley},
  {Valenti}, {Buonanno}, {Calamida}, {Castellani}, {da Silva},
  {Degl'Innocenti}, {Di Cecco}, {Fabrizio}, {Freedman}, {Giuffrida}, {Lub},
  {Madore}, {Marconi}, {Marinoni}, {Matsunaga}, {Monelli}, {Persson},
  {Piersimoni}, {Pietrinferni}, {Prada-Moroni}, {Pulone}, {Stellingwerf},
  {Tognelli}, \& {Walker}}]{braga16}
{Braga}, V.~F., {Stetson}, P.~B., {Bono}, G., {et~al.} 2016, \aj, 152, 170

\bibitem[{{Brown} {et~al.}(2000){Brown}, {Bowers}, {Kimble}, {Sweigart}, \&
  {Ferguson}}]{brown00}
{Brown}, T.~M., {Bowers}, C.~W., {Kimble}, R.~A., {Sweigart}, A.~V., \&
  {Ferguson}, H.~C. 2000, \apj, 532, 308

\bibitem[{{Buchler} \& {Goupil}(1988)}]{buchler88}
{Buchler}, J.~R. \& {Goupil}, M.-J. 1988, \aap, 190, 137

\bibitem[{{Calamida} {et~al.}(2017){Calamida}, {Strampelli}, {Rest}, {Bono},
  {Ferraro}, {Saha}, {Iannicola}, {Scolnic}, {James}, {Smith}, \&
  {Zenteno}}]{calamida2017}
{Calamida}, A., {Strampelli}, G., {Rest}, A., {et~al.} 2017, \aj, 153, 175

\bibitem[{{Carretta} {et~al.}(2009){Carretta}, {Bragaglia}, {Gratton},
  {D'Orazi}, \& {Lucatello}}]{carretta09}
{Carretta}, E., {Bragaglia}, A., {Gratton}, R., {D'Orazi}, V., \& {Lucatello},
  S. 2009, \aap, 508, 695

\bibitem[{{Cassisi} \& {Salaris}(2011)}]{cassisi11}
{Cassisi}, S. \& {Salaris}, M. 2011, \apjl, 728, L43

\bibitem[{{Cassisi} \& {Salaris}(2013)}]{cassisi13}
{Cassisi}, S. \& {Salaris}, M. 2013, {Old Stellar Populations: How to Study the
  Fossil Record of Galaxy Formation}

\bibitem[{{Cassisi} {et~al.}(2003){Cassisi}, {Schlattl}, {Salaris}, \&
  {Weiss}}]{cassisi03}
{Cassisi}, S., {Schlattl}, H., {Salaris}, M., \& {Weiss}, A. 2003, \apjl, 582,
  L43

\bibitem[{{Castellani} {et~al.}(2003){Castellani}, {Caputo}, \&
  {Castellani}}]{castellani2003}
{Castellani}, M., {Caputo}, F., \& {Castellani}, V. 2003, \aap, 410, 871

\bibitem[{{Castellani} \& {Castellani}(1993)}]{castellani93}
{Castellani}, M. \& {Castellani}, V. 1993, \apj, 407, 649

\bibitem[{{Castellani} {et~al.}(2006){Castellani}, {Castellani}, \& {Prada
  Moroni}}]{castellani06}
{Castellani}, M., {Castellani}, V., \& {Prada Moroni}, P.~G. 2006, \aap, 457,
  569

\bibitem[{{Castellani} {et~al.}(2007){Castellani}, {Calamida}, {Bono},
  {Stetson}, {Freyhammer}, {Degl'Innocenti}, {Moroni}, {Monelli}, {Corsi},
  {Nonino}, {Buonanno}, {Caputo}, {Castellani}, {Dall'Ora}, {Del Principe},
  {Ferraro}, {Iannicola}, {Piersimoni}, {Pulone}, \& {Vuerli}}]{castellani2007}
{Castellani}, V., {Calamida}, A., {Bono}, G., {et~al.} 2007, \apj, 663, 1021

\bibitem[{{Castellani} {et~al.}(1991){Castellani}, {Chieffi}, \&
  {Pulone}}]{castellani91}
{Castellani}, V., {Chieffi}, A., \& {Pulone}, L. 1991, \apjs, 76, 911

\bibitem[{{Castelli} {et~al.}(1997){Castelli}, {Gratton}, \&
  {Kurucz}}]{castelli97}
{Castelli}, F., {Gratton}, R.~G., \& {Kurucz}, R.~L. 1997, \aap, 318, 841

\bibitem[{{Castelli} \& {Kurucz}(2003)}]{castelli03}
{Castelli}, F. \& {Kurucz}, R.~L. 2003, in IAU Symposium, Vol. 210, Modelling
  of Stellar Atmospheres, ed. N.~{Piskunov}, W.~W. {Weiss}, \& D.~F. {Gray},
  A20

\bibitem[{{Chiosi} {et~al.}(1993){Chiosi}, {Wood}, \& {Capitanio}}]{chiosi93}
{Chiosi}, C., {Wood}, P.~R., \& {Capitanio}, N. 1993, \apjs, 86, 541

\bibitem[{{Clement} {et~al.}(2001){Clement}, {Muzzin}, {Dufton}, {Ponnampalam},
  {Wang}, {Burford}, {Richardson}, {Rosebery}, {Rowe}, \& {Hogg}}]{clement01}
{Clement}, C.~M., {Muzzin}, A., {Dufton}, Q., {et~al.} 2001, \aj, 122, 2587

\bibitem[{{Constantino} {et~al.}(2016){Constantino}, {Campbell}, {Lattanzio},
  \& {van Duijneveldt}}]{constantino16}
{Constantino}, T., {Campbell}, S.~W., {Lattanzio}, J.~C., \& {van Duijneveldt},
  A. 2016, \mnras, 456, 3866

\bibitem[{{Cristallo} {et~al.}(2009){Cristallo}, {Straniero}, {Gallino},
  {Piersanti}, {Dom{\'{\i}}nguez}, \& {Lederer}}]{cristallo09}
{Cristallo}, S., {Straniero}, O., {Gallino}, R., {et~al.} 2009, \apj, 696, 797

\bibitem[{{Dalessandro} {et~al.}(2016){Dalessandro}, {Lapenna}, {Mucciarelli},
  {Origlia}, {Ferraro}, \& {Lanzoni}}]{dalessandro2016}
{Dalessandro}, E., {Lapenna}, E., {Mucciarelli}, A., {et~al.} 2016, \apj, 829,
  77

\bibitem[{{Dalessandro} {et~al.}(2013){Dalessandro}, {Salaris}, {Ferraro},
  {Mucciarelli}, \& {Cassisi}}]{dalessandro13a}
{Dalessandro}, E., {Salaris}, M., {Ferraro}, F.~R., {Mucciarelli}, A., \&
  {Cassisi}, S. 2013, \mnras, 430, 459

\bibitem[{{D'Cruz} {et~al.}(1996){D'Cruz}, {Dorman}, {Rood}, \&
  {O'Connell}}]{d'cruz96}
{D'Cruz}, N.~L., {Dorman}, B., {Rood}, R.~T., \& {O'Connell}, R.~W. 1996, \apj,
  466, 359

\bibitem[{{Di Criscienzo} {et~al.}(2007){Di Criscienzo}, {Caputo}, {Marconi},
  \& {Cassisi}}]{dicriscienzo07}
{Di Criscienzo}, M., {Caputo}, F., {Marconi}, M., \& {Cassisi}, S. 2007, \aap,
  471, 893

\bibitem[{{Dorman} \& {Rood}(1993)}]{dorman93}
{Dorman}, B. \& {Rood}, R.~T. 1993, \apj, 409, 387

\bibitem[{{Dotter}(2008)}]{dotter08}
{Dotter}, A. 2008, \apjl, 687, L21

\bibitem[{{Dziembowski} \& {Cassisi}(1999)}]{dziembowski99}
{Dziembowski}, W.~A. \& {Cassisi}, S. 1999, \actaa, 49, 371

\bibitem[{{Fabrizio} {et~al.}(2019){Fabrizio}, {Bono}, {Braga}, {Magurno},
  {Marinoni}, {Marrese}, {Ferraro}, {Fiorentino}, {Giuffrida}, {Iannicola},
  {Monelli}, {Altavilla}, {Chaboyer}, {Dall'Ora}, {Gilligan}, {Layden},
  {Marengo}, {Nonino}, {Preston}, {Sesar}, {Sneden}, {Valenti}, {Th{\'e}venin},
  \& {Zoccali}}]{fabrizio2019}
{Fabrizio}, M., {Bono}, G., {Braga}, V.~F., {et~al.} 2019, \apj, 882, 169

\bibitem[{{Feast} {et~al.}(2008){Feast}, {Laney}, {Kinman}, {van Leeuwen}, \&
  {Whitelock}}]{feast08}
{Feast}, M.~W., {Laney}, C.~D., {Kinman}, T.~D., {van Leeuwen}, F., \&
  {Whitelock}, P.~A. 2008, \mnras, 386, 2115

\bibitem[{{Ferguson} {et~al.}(2005){Ferguson}, {Alexander}, {Allard}, {Barman},
  {Bodnarik}, {Hauschildt}, {Heffner-Wong}, \& {Tamanai}}]{ferguson05}
{Ferguson}, J.~W., {Alexander}, D.~R., {Allard}, F., {et~al.} 2005, \apj, 623,
  585

\bibitem[{{Fiorentino} {et~al.}(2017){Fiorentino}, {Bellazzini}, {Ciliegi},
  {Chauvin}, {Dout{'e}}, {D'Orazi}, {Maiorano}, {Mannucci}, {Mapelli}, {Podio},
  {Saracco}, \& {Spavone}}]{fiorentino17}
{Fiorentino}, G., {Bellazzini}, M., {Ciliegi}, P., {et~al.} 2017, arXiv
  e-prints, arXiv:1712.04222

\bibitem[{{Fricke} {et~al.}(1971){Fricke}, {Stobie}, \&
  {Strittmatter}}]{fricke71}
{Fricke}, K., {Stobie}, R.~S., \& {Strittmatter}, P.~A. 1971, \mnras, 154, 23

\bibitem[{{Giannone} \& {Rossi}(1977)}]{giannone77}
{Giannone}, P. \& {Rossi}, L. 1977, \memsai, 48, 776

\bibitem[{{Gingold}(1974)}]{gingold74}
{Gingold}, R.~A. 1974, \apj, 193, 177

\bibitem[{{Gingold}(1976)}]{gingold76}
{Gingold}, R.~A. 1976, \apj, 204, 116

\bibitem[{{Gingold}(1985)}]{gingold85}
{Gingold}, R.~A. 1985, \memsai, 56, 169

\bibitem[{{Gonzalez} \& {Lambert}(1997)}]{gonzalez1997b}
{Gonzalez}, G. \& {Lambert}, D.~L. 1997, \aj, 114, 341

\bibitem[{{Gonzalez} {et~al.}(1997){Gonzalez}, {Lambert}, \&
  {Giridhar}}]{gonzalez1997a}
{Gonzalez}, G., {Lambert}, D.~L., \& {Giridhar}, S. 1997, \apj, 479, 427

\bibitem[{{Gonzalez} \& {Wallerstein}(1994)}]{gonzalez1994b}
{Gonzalez}, G. \& {Wallerstein}, G. 1994, \aj, 108, 1325

\bibitem[{{Gonzalez} \& {Wallerstein}(1996)}]{gonzalez96}
{Gonzalez}, G. \& {Wallerstein}, G. 1996, \mnras, 280, 515

\bibitem[{{Gratton} {et~al.}(2004{\natexlab{a}}){Gratton}, {Sneden}, \&
  {Carretta}}]{gratton04}
{Gratton}, R., {Sneden}, C., \& {Carretta}, E. 2004{\natexlab{a}}, \araa, 42,
  385

\bibitem[{{Gratton} {et~al.}(2004{\natexlab{b}}){Gratton}, {Bragaglia},
  {Clementini}, {Carretta}, {Di Fabrizio}, {Maio}, \& {Taribello}}]{gratton04a}
{Gratton}, R.~G., {Bragaglia}, A., {Clementini}, G., {et~al.}
  2004{\natexlab{b}}, \aap, 421, 937

\bibitem[{{Gratton} {et~al.}(2003){Gratton}, {Carretta}, {Claudi}, {Lucatello},
  \& {Barbieri}}]{gratton2003}
{Gratton}, R.~G., {Carretta}, E., {Claudi}, R., {Lucatello}, S., \& {Barbieri},
  M. 2003, \aap, 404, 187

\bibitem[{{Greggio} \& {Renzini}(1990)}]{greggio90}
{Greggio}, L. \& {Renzini}, A. 1990, \apj, 364, 35

\bibitem[{{Groenewegen} \&
  {Jurkovic}(2017{\natexlab{a}})}]{groenewegenjurkovic2017}
{Groenewegen}, M.~A.~T. \& {Jurkovic}, M.~I. 2017{\natexlab{a}}, \aap, 603, A70

\bibitem[{{Groenewegen} \&
  {Jurkovic}(2017{\natexlab{b}})}]{groenewegenjurkovic2017b}
{Groenewegen}, M.~A.~T. \& {Jurkovic}, M.~I. 2017{\natexlab{b}}, \aap, 604, A29

\bibitem[{{Gustafsson} {et~al.}(1975){Gustafsson}, {Bell}, {Eriksson}, \&
  {Nordlund}}]{gustafsson75}
{Gustafsson}, B., {Bell}, R.~A., {Eriksson}, K., \& {Nordlund}, A. 1975, \aap,
  42, 407

\bibitem[{{Gustafsson} {et~al.}(2008){Gustafsson}, {Edvardsson}, {Eriksson},
  {J{\o}rgensen}, {Nordlund}, \& {Plez}}]{gustafsson08}
{Gustafsson}, B., {Edvardsson}, B., {Eriksson}, K., {et~al.} 2008, \aap, 486,
  951

\bibitem[{{Guzik} {et~al.}(2000){Guzik}, {Kaye}, {Bradley}, {Cox}, \&
  {Neuforge}}]{guzik00}
{Guzik}, J.~A., {Kaye}, A.~B., {Bradley}, P.~A., {Cox}, A.~N., \& {Neuforge},
  C. 2000, \apjl, 542, L57

\bibitem[{{Harris}(1981)}]{harris81}
{Harris}, H.~C. 1981, \aj, 86, 719

\bibitem[{{Harris} \& {Wallerstein}(1984)}]{harris_wallerstein1984}
{Harris}, H.~C. \& {Wallerstein}, G. 1984, \aj, 89, 379

\bibitem[{{Harris}(1996)}]{harris96}
{Harris}, W.~E. 1996, \aj, 112, 1487

\bibitem[{{Heber} {et~al.}(1986){Heber}, {Kudritzki}, {Caloi}, {Castellani}, \&
  {Danziger}}]{kudritzki86}
{Heber}, U., {Kudritzki}, R.~P., {Caloi}, V., {Castellani}, V., \& {Danziger},
  J. 1986, \aap, 162, 171

\bibitem[{{Iben} \& {Huchra}(1970)}]{iben70}
{Iben}, Jr., I. \& {Huchra}, J. 1970, \apjl, 162, L43

\bibitem[{{Iben} \& {Truran}(1978)}]{iben78}
{Iben}, Jr., I. \& {Truran}, J.~W. 1978, \apj, 220, 980

\bibitem[{{Iglesias} \& {Rogers}(1996)}]{iglesias96}
{Iglesias}, C.~A. \& {Rogers}, F.~J. 1996, \apj, 464, 943

\bibitem[{{Iwanek} {et~al.}(2018){Iwanek}, {Soszy{\'n}ski}, {Skowron},
  {Skowron}, {Mr{\'o}z}, {Koz{\l}owski}, {Udalski}, {Szyma{\'n}ski},
  {Pietrukowicz}, {Poleski}, \& {Jacyszyn-Dobrzeniecka}}]{iwanek2018}
{Iwanek}, P., {Soszy{\'n}ski}, I., {Skowron}, D., {et~al.} 2018, \actaa, 68,
  213

\bibitem[{{Johnson} \& {Pilachowski}(2010)}]{johnson10}
{Johnson}, C.~I. \& {Pilachowski}, C.~A. 2010, \apj, 722, 1373

\bibitem[{{Kippenhahn} {et~al.}(1967){Kippenhahn}, {Weigert}, \&
  {Hofmeister}}]{kippenhahn67}
{Kippenhahn}, R., {Weigert}, A., \& {Hofmeister}, E. 1967, Methods in
  Computational Physics, 7, 129

\bibitem[{{Kovacs} \& {Buchler}(1993)}]{kovacs93}
{Kovacs}, G. \& {Buchler}, J.~R. 1993, \apj, 404, 765

\bibitem[{{Kovtyukh} {et~al.}(2018){Kovtyukh}, {Wallerstein}, {Yegorova},
  {Andrievsky}, {Korotin}, {Saviane}, {Belik}, {Davis}, \&
  {Farrell}}]{kovtyukh2018a}
{Kovtyukh}, V., {Wallerstein}, G., {Yegorova}, I., {et~al.} 2018, \pasp, 130,
  054201

\bibitem[{{Kurucz}(1979)}]{kurucz79}
{Kurucz}, R.~L. 1979, \apjs, 40, 1

\bibitem[{{Latour} {et~al.}(2014){Latour}, {Randall}, {Fontaine}, {Bono},
  {Calamida}, \& {Brassard}}]{latour14}
{Latour}, M., {Randall}, S.~K., {Fontaine}, G., {et~al.} 2014, \apj, 795, 106

\bibitem[{{Lebzelter} \& {Wood}(2016)}]{lebzelter2016}
{Lebzelter}, T. \& {Wood}, P.~R. 2016, \aap, 585, A111

\bibitem[{{Lee} {et~al.}(1990){Lee}, {Demarque}, \& {Zinn}}]{lee90}
{Lee}, Y.-W., {Demarque}, P., \& {Zinn}, R. 1990, \apj, 350, 155

\bibitem[{{Lemasle} {et~al.}(2015){Lemasle}, {Kovtyukh}, {Bono}, {Fran{\c
  c}ois}, {Saviane}, {Yegorova}, {Genovali}, {Inno}, {Galazutdinov}, \& {da
  Silva}}]{lemasle15}
{Lemasle}, B., {Kovtyukh}, V., {Bono}, G., {et~al.} 2015, \aap, 579, A47

\bibitem[{{Luck} \& {Bond}(1989)}]{luck1989}
{Luck}, R.~E. \& {Bond}, H.~E. 1989, \apj, 342, 476

\bibitem[{{Maas} {et~al.}(2007){Maas}, {Giridhar}, \& {Lambert}}]{maas2007}
{Maas}, T., {Giridhar}, S., \& {Lambert}, D.~L. 2007, \apj, 666, 378

\bibitem[{{Madore}(1982)}]{madore82}
{Madore}, B.~F. 1982, \apj, 253, 575

\bibitem[{{Madore} {et~al.}(2017){Madore}, {Freedman}, \& {Moak}}]{madore17}
{Madore}, B.~F., {Freedman}, W.~L., \& {Moak}, S. 2017, \apj, 842, 42

\bibitem[{{Magurno} {et~al.}(2019){Magurno}, {Sneden}, {Bono}, {Braga},
  {Mateo}, {Persson}, {Preston}, {Th{\'e}venin}, {da Silva}, {Dall'Ora},
  {Fabrizio}, {Ferraro}, {Fiorentino}, {Iannicola}, {Inno}, {Marengo},
  {Marinoni}, {Marrese}, {Mart{\'\i}nez-V{\'a}zquez}, {Matsunaga}, {Monelli},
  {Neeley}, {Nonino}, \& {Walker}}]{magurno2019}
{Magurno}, D., {Sneden}, C., {Bono}, G., {et~al.} 2019, \apj, 881, 104

\bibitem[{{Magurno} {et~al.}(2018){Magurno}, {Sneden}, {Braga}, {Bono},
  {Mateo}, {Persson}, {Dall'Ora}, {Marengo}, {Monelli}, \&
  {Neeley}}]{magurno2018}
{Magurno}, D., {Sneden}, C., {Braga}, V.~F., {et~al.} 2018, \apj, 864, 57

\bibitem[{{Marconi} {et~al.}(2018){Marconi}, {Bono}, {Pietrinferni}, {Braga},
  {Castellani}, \& {Stellingwerf}}]{marconi2018}
{Marconi}, M., {Bono}, G., {Pietrinferni}, A., {et~al.} 2018, \apjl, 864, L13

\bibitem[{{Marconi} {et~al.}(2015){Marconi}, {Coppola}, {Bono}, {Braga},
  {Pietrinferni}, {Buonanno}, {Castellani}, {Musella}, {Ripepi}, \&
  {Stellingwerf}}]{marconi15}
{Marconi}, M., {Coppola}, G., {Bono}, G., {et~al.} 2015, \apj, 808, 50

\bibitem[{{Marconi} \& {Di Criscienzo}(2007)}]{marconi07}
{Marconi}, M. \& {Di Criscienzo}, M. 2007, \aap, 467, 223

\bibitem[{{Mart{\'{\i}}nez-V{\'a}zquez}
  {et~al.}(2016){Mart{\'{\i}}nez-V{\'a}zquez}, {Stetson}, {Monelli}, {Bernard},
  {Fiorentino}, {Gallart}, {Bono}, {Cassisi}, {Dall'Ora}, {Ferraro},
  {Iannicola}, \& {Walker}}]{martinezvazquez16b}
{Mart{\'{\i}}nez-V{\'a}zquez}, C.~E., {Stetson}, P.~B., {Monelli}, M., {et~al.}
  2016, \mnras, 462, 4349

\bibitem[{{Matsunaga} {et~al.}(2013){Matsunaga}, {Feast}, {Kawadu},
  {Nishiyama}, {Nagayama}, {Nagata}, {Tamura}, {Bono}, \&
  {Kobayashi}}]{matsunaga2013}
{Matsunaga}, N., {Feast}, M.~W., {Kawadu}, T., {et~al.} 2013, \mnras, 429, 385

\bibitem[{{Matsunaga} {et~al.}(2006){Matsunaga}, {Fukushi}, {Nakada},
  {Tanab{\'e}}, {Feast}, {Menzies}, {Ita}, {Nishiyama}, {Baba}, {Naoi},
  {Nakaya}, {Kawadu}, {Ishihara}, \& {Kato}}]{matsunaga06}
{Matsunaga}, N., {Fukushi}, H., {Nakada}, Y., {et~al.} 2006, \mnras, 370, 1979

\bibitem[{{Miller Bertolami} {et~al.}(2008){Miller Bertolami}, {Althaus},
  {Unglaub}, \& {Weiss}}]{millerbertolami08}
{Miller Bertolami}, M.~M., {Althaus}, L.~G., {Unglaub}, K., \& {Weiss}, A.
  2008, \aap, 491, 253

\bibitem[{{Minniti}(1995)}]{minniti1995}
{Minniti}, D. 1995, \aap, 303, 468

\bibitem[{{Moehler} {et~al.}(2011){Moehler}, {Dreizler}, {Lanz}, {Bono},
  {Sweigart}, {Calamida}, \& {Nonino}}]{moehler11}
{Moehler}, S., {Dreizler}, S., {Lanz}, T., {et~al.} 2011, \aap, 526, A136

\bibitem[{{Origlia} {et~al.}(2014){Origlia}, {Ferraro}, {Fabbri}, {Fusi Pecci},
  {Dalessandro}, {Rich}, \& {Valenti}}]{origlia14}
{Origlia}, L., {Ferraro}, F.~R., {Fabbri}, S., {et~al.} 2014, \aap, 564, A136

\bibitem[{{Percy} {et~al.}(1991){Percy}, {Sasselov}, {Alfred}, \&
  {Scott}}]{percy1991}
{Percy}, J.~R., {Sasselov}, D.~D., {Alfred}, A., \& {Scott}, G. 1991, \apj,
  375, 691

\bibitem[{{Piatti} \& {Koch}(2018)}]{piatti2018}
{Piatti}, A.~E. \& {Koch}, A. 2018, \apj, 867, 8

\bibitem[{{Pietrinferni} {et~al.}(2006{\natexlab{a}}){Pietrinferni}, {Cassisi},
  {Bono}, {Stetson}, {Iannicola}, {Castellani}, {Buonanno}, \&
  {Zoccali}}]{pietrinferni06a}
{Pietrinferni}, A., {Cassisi}, S., {Bono}, G., {et~al.} 2006{\natexlab{a}},
  \memsai, 77, 144

\bibitem[{{Pietrinferni} {et~al.}(2006{\natexlab{b}}){Pietrinferni}, {Cassisi},
  {Salaris}, \& {Castelli}}]{pietrinferni06b}
{Pietrinferni}, A., {Cassisi}, S., {Salaris}, M., \& {Castelli}, F.
  2006{\natexlab{b}}, \apj, 642, 797

\bibitem[{{Pilecki} {et~al.}(2018){Pilecki}, {Dervi{\textcommabelow
  s}o{\u{g}}lu}, {Gieren}, {Smolec}, {Soszy{\'n}ski}, {Pietrzy{\'n}ski},
  {Thompson}, \& {Taormina}}]{pilecki2018}
{Pilecki}, B., {Dervi{\textcommabelow s}o{\u{g}}lu}, A., {Gieren}, W., {et~al.}
  2018, \apj, 868, 30

\bibitem[{{Pilecki} {et~al.}(2017){Pilecki}, {Gieren}, {Smolec},
  {Pietrzy{\'n}ski}, {Thompson}, {Anderson}, {Bono}, {Soszy{\'n}ski},
  {Kervella}, {Nardetto}, {Taormina}, {St{\c e}pie{\'n}}, \&
  {Wielg{\'o}rski}}]{pilecki17}
{Pilecki}, B., {Gieren}, W., {Smolec}, R., {et~al.} 2017, \apj, 842, 110

\bibitem[{{Pritzl} {et~al.}(2002){Pritzl}, {Smith}, {Catelan}, \&
  {Sweigart}}]{pritzl2002a}
{Pritzl}, B.~J., {Smith}, H.~A., {Catelan}, M., \& {Sweigart}, A.~V. 2002, \aj,
  124, 949

\bibitem[{{Pritzl} {et~al.}(2003){Pritzl}, {Smith}, {Stetson}, {Catelan},
  {Sweigart}, {Layden}, \& {Rich}}]{pritzl03}
{Pritzl}, B.~J., {Smith}, H.~A., {Stetson}, P.~B., {et~al.} 2003, \aj, 126,
  1381

\bibitem[{{Rabidoux} {et~al.}(2010){Rabidoux}, {Smith}, {Pritzl}, {Osborn},
  {Kuehn}, {Randall}, {Lustig}, {Wells}, {Taylor}, {De Lee}, {Kinemuchi},
  {LaCluyz{\'e}}, {Hartley}, {Greenwood}, {Ingber}, {Ireland}, {Pellegrini},
  {Anderson}, {Purdum}, {Lacy}, {Curtis}, {Smolinski}, \&
  {Danford}}]{rabidoux10}
{Rabidoux}, K., {Smith}, H.~A., {Pritzl}, B.~J., {et~al.} 2010, \aj, 139, 2300

\bibitem[{{Rich} {et~al.}(2012){Rich}, {Origlia}, \& {Valenti}}]{rich2012}
{Rich}, R.~M., {Origlia}, L., \& {Valenti}, E. 2012, \apj, 746, 59

\bibitem[{{Riess} {et~al.}(2019){Riess}, {Casertano}, {Yuan}, {Macri}, \&
  {Scolnic}}]{riess2019}
{Riess}, A.~G., {Casertano}, S., {Yuan}, W., {Macri}, L.~M., \& {Scolnic}, D.
  2019, \apj, 876, 85

\bibitem[{{Ripepi} {et~al.}(2019){Ripepi}, {Molinaro}, {Musella}, {Marconi},
  {Leccia}, \& {Eyer}}]{ripepi2019}
{Ripepi}, V., {Molinaro}, R., {Musella}, I., {et~al.} 2019, \aap, 625, A14

\bibitem[{{Ripepi} {et~al.}(2015){Ripepi}, {Moretti}, {Marconi}, {Clementini},
  {Cioni}, {de Grijs}, {Emerson}, {Groenewegen}, {Ivanov}, {Muraveva},
  {Piatti}, \& {Subramanian}}]{ripepi2015}
{Ripepi}, V., {Moretti}, M.~I., {Marconi}, M., {et~al.} 2015, \mnras, 446, 3034

\bibitem[{{Rodgers} \& {Bell}(1968)}]{rodgers1968}
{Rodgers}, A.~W. \& {Bell}, R.~A. 1968, \mnras, 139, 175

\bibitem[{{Salaris} {et~al.}(2013){Salaris}, {Althaus}, \&
  {Garc{\'{\i}}a-Berro}}]{salaris13b}
{Salaris}, M., {Althaus}, L.~G., \& {Garc{\'{\i}}a-Berro}, E. 2013, \aap, 555,
  A96

\bibitem[{{Salaris} \& {Cassisi}(2005)}]{salaris05}
{Salaris}, M. \& {Cassisi}, S. 2005, {Evolution of Stars and Stellar
  Populations} (Chichester: John Wiley \& Sons, Ltd)

\bibitem[{{Salaris} {et~al.}(2008){Salaris}, {Cassisi}, \&
  {Pietrinferni}}]{salaris08}
{Salaris}, M., {Cassisi}, S., \& {Pietrinferni}, A. 2008, \apjl, 678, L25

\bibitem[{{Savino} {et~al.}(2015){Savino}, {Salaris}, \& {Tolstoy}}]{savino15}
{Savino}, A., {Salaris}, M., \& {Tolstoy}, E. 2015, \aap, 583, A126

\bibitem[{{Seaton}(2007)}]{seaton07}
{Seaton}, M.~J. 2007, \mnras, 382, 245

\bibitem[{{Soszy{\'n}ski} {et~al.}(2019){Soszy{\'n}ski}, {Smolec}, {Udalski},
  \& {Pietrukowicz}}]{soszynski2019b}
{Soszy{\'n}ski}, I., {Smolec}, R., {Udalski}, A., \& {Pietrukowicz}, P. 2019,
  \apj, 873, 43

\bibitem[{{Soszy{\'n}ski} {et~al.}(2011){Soszy{\'n}ski}, {Udalski},
  {Pietrukowicz}, {Szyma{\'n}ski}, {Kubiak}, {Pietrzy{\'n}ski}, {Wyrzykowski},
  {Ulaczyk}, {Poleski}, \& {Koz{\l}owski}}]{soszynski2011}
{Soszy{\'n}ski}, I., {Udalski}, A., {Pietrukowicz}, P., {et~al.} 2011, \actaa,
  61, 285

\bibitem[{{Soszy{\'n}ski} {et~al.}(2008){Soszy{\'n}ski}, {Udalski},
  {Szyma{\'n}ski}, {Kubiak}, {Pietrzy{\'n}ski}, {Wyrzykowski}, {Szewczyk},
  {Ulaczyk}, \& {Poleski}}]{soszynski08c}
{Soszy{\'n}ski}, I., {Udalski}, A., {Szyma{\'n}ski}, M.~K., {et~al.} 2008,
  \actaa, 58, 293

\bibitem[{{Soszy{\'n}ski} {et~al.}(2018){Soszy{\'n}ski}, {Udalski},
  {Szyma{\'n}ski}, {Wyrzykowski}, {Ulaczyk}, {Poleski}, {Pietrukowicz},
  {Koz{\l}owski}, {Skowron}, {Skowron}, {Mr{\'o}z}, {Rybicki}, \&
  {Iwanek}}]{soszynski2018}
{Soszy{\'n}ski}, I., {Udalski}, A., {Szyma{\'n}ski}, M.~K., {et~al.} 2018,
  \actaa, 68, 89

\bibitem[{{Soszy{\'n}ski} {et~al.}(2017){Soszy{\'n}ski}, {Udalski},
  {Szyma{\'n}ski}, {Wyrzykowski}, {Ulaczyk}, {Poleski}, {Pietrukowicz},
  {Koz{\l}owski}, {Skowron}, {Skowron}, {Mr{\'o}z}, {Pawlak}, {Rybicki}, \&
  {Jacyszyn-Dobrzeniecka}}]{soszynski2017}
{Soszy{\'n}ski}, I., {Udalski}, A., {Szyma{\'n}ski}, M.~K., {et~al.} 2017,
  \actaa, 67, 297

\bibitem[{{Tammann} {et~al.}(2003){Tammann}, {Sandage}, \&
  {Reindl}}]{tammann03}
{Tammann}, G.~A., {Sandage}, A., \& {Reindl}, B. 2003, \aap, 404, 423

\bibitem[{{Torelli} {et~al.}(2019){Torelli}, {Iannicola}, {Stetson}, {Ferraro},
  {Bono}, {Salaris}, {Castellani}, {Dall'Ora}, {Fontana}, {Monelli}, \&
  {Pietrinferni}}]{torelli2019}
{Torelli}, M., {Iannicola}, G., {Stetson}, P.~B., {et~al.} 2019, \aap, 629, A53

\bibitem[{{Tuggle} \& {Iben}(1973)}]{tuggle73}
{Tuggle}, R.~S. \& {Iben}, Jr., I. 1973, \apj, 186, 593

\bibitem[{{Udalski} {et~al.}(2015){Udalski}, {Szyma{\'n}ski}, \&
  {Szyma{\'n}ski}}]{udalski2015}
{Udalski}, A., {Szyma{\'n}ski}, M.~K., \& {Szyma{\'n}ski}, G. 2015, \actaa, 65,
  1

\bibitem[{{van Albada} \& {Baker}(1973)}]{vanalbada73}
{van Albada}, T.~S. \& {Baker}, N. 1973, \apj, 185, 477

\bibitem[{{VandenBerg} {et~al.}(2013){VandenBerg}, {Brogaard}, {Leaman}, \&
  {Casagrande}}]{vandenberg13}
{VandenBerg}, D.~A., {Brogaard}, K., {Leaman}, R., \& {Casagrande}, L. 2013,
  \apj, 775, 134

\bibitem[{{Vassiliadis} \& {Wood}(1993)}]{vassiliadis93}
{Vassiliadis}, E. \& {Wood}, P.~R. 1993, \apj, 413, 641

\bibitem[{{Wagenhuber} \& {Groenewegen}(1998)}]{wagenhuber98}
{Wagenhuber}, J. \& {Groenewegen}, M.~A.~T. 1998, \aap, 340, 183

\bibitem[{{Walker} \& {Terndrup}(1991)}]{walker1991a}
{Walker}, A.~R. \& {Terndrup}, D.~M. 1991, \apj, 378, 119

\bibitem[{{Wallerstein}(2002)}]{wallerstein2002}
{Wallerstein}, G. 2002, \pasp, 114, 689

\bibitem[{{Wallerstein} \& {Farrell}(2018)}]{wallerstein_farrell2018}
{Wallerstein}, G. \& {Farrell}, E.~M. 2018, \aj, 156, 299

\bibitem[{{Wallerstein} \& {Gonzalez}(1996)}]{wallerstein1996}
{Wallerstein}, G. \& {Gonzalez}, G. 1996, \mnras, 282, 1236

\bibitem[{{Wallerstein} {et~al.}(2008){Wallerstein}, {Kovtyukh}, \&
  {Andrievsky}}]{wallerstein2008}
{Wallerstein}, G., {Kovtyukh}, V.~V., \& {Andrievsky}, S.~M. 2008, \pasp, 120,
  361

\bibitem[{{Wallerstein} {et~al.}(2000){Wallerstein}, {Matt}, \&
  {Gonzalez}}]{wallerstein2000}
{Wallerstein}, G., {Matt}, S., \& {Gonzalez}, G. 2000, \mnras, 311, 414

\bibitem[{{Weiss} \& {Ferguson}(2009)}]{weiss09}
{Weiss}, A. \& {Ferguson}, J.~W. 2009, \aap, 508, 1343

\bibitem[{{Zoccali} {et~al.}(2017){Zoccali}, {Vasquez}, {Gonzalez}, {Valenti},
  {Rojas-Arriagada}, {Minniti}, {Rejkuba}, {Minniti}, {McWilliam}, {Babusiaux},
  {Hill}, \& {Renzini}}]{zoccali2017}
{Zoccali}, M., {Vasquez}, S., {Gonzalez}, O.~A., {et~al.} 2017, A\&A, 599, A12

\end{thebibliography}

%%%%%%%%%%%%%%%%%%%%%%%%%%%%%%%%%%%%%%%%%%%%%%%%%%     
     
%%%%%%%%%%%%%%%%% APPENDICES %%%%%%%%%%%%%%%%%%%%%     
     
\begin{appendix}  
%______________________________________________________________________________     
\section{The metallicity distribution of Type II Cepheids}\label{sec:metallicity}     
     
The metallicity distribution of TIICs is quite broad.      
Cluster TIICs have been identified in metal-poor globulars like      
NGC~7078 ([Fe/H]$\sim$--2.4), including two TIICs       
but no RRLs, metal-intermediate globulars like M13 ([Fe/H]$\sim$--1.5) \citep{clement01},      
in more metal-rich ones like NGC~6441 (${\rm [Fe/H]}\sim$--0.5),   
NGC~6388 \citep[${\rm [Fe/H]}\sim$--0.6,][]{pritzl03}      
and $\omega$ Cen, characterized by a very broad      
metallicity distribution \citep{johnson10,calamida2017}.      
The top panel of Fig.~\ref{fig:metallicity} shows the metallicity distribution of TIICs in globulars      
according to the online catalog by \cite{clement01}.      
Cluster metallicities (see Table~\ref{tab:cluster_met})       
are based on the metallicity scale by \cite{carretta09}. Note that the two most metal-rich      
clusters (NGC~6388, NGC~6441) hosting TIICs cause an isolated secondary peak in the      
metallicity distribution (\citealt{pritzl2002a,pritzl03}, Fig.~\ref{fig:metallicity}). It is not clear yet     
whether this clumpy distribution is intrinsic or affected by an      
observational bias, because we still lack quantitative analyses based on      
proper motions, radial velocities and distances, concerning the presence of TIICs in metal-rich      
clusters of the Galactic bulge.

%%%%%%%%%%%%%%%%%%%%%%%%%%%%%%%%%%%%%%%%%%%%%%%%%%%%%%%%%%%%%%%%%%%%%%%%5  
% Table A-1 
%%%%%%%%%%%%%%%%%%%%%%%%%%%%%%%%%%%%%%%%%%%%%%%%%%%%%%%%%%%%%%%%%%%%%%%%5  
\begin{table}     
\scriptsize     
\caption{Iron abundances for Galactic globular cluster TIICs.}     
\label{tab:cluster_met}     
\centering     
\scriptsize     
\begin{tabular}{llrr|llrr}     
\hline     
\hline     
Cluster &  ID       & [Fe/H] & n &            Cluster  &  ID   &   [Fe/H] & n  \\     
\hline     
      HP 1 &        & --1.57        &  2 &    NGC 6341 & M92   & --2.35 &  1 \\     
  NGC 2419 &        & --2.20        &  1 &    NGC 6388 &       & --0.45 & 12 \\     
  NGC 2808 &        & --1.18        &  1 &    NGC 6401 &       & --1.01 &  1 \\     
  NGC 5139 &  \wcen & --1.61/--1.95 &  7 &    NGC 6402 & M14   & --1.39 &  5 \\     
  NGC 5272 &  M3    & --1.50        &  1 &    NGC 6441 &       & --0.44 &  8 \\     
  NGC 5286 &        & --1.70        &  1 &    NGC 6453 &       & --1.48 &  2 \\     
  NGC 5904 &  M5    & --1.33        &  2 &    NGC 6522 &       & --1.45 &  2 \\     
  NGC 5986 &        & --1.63        &  1 &    NGC 6626 & M28   & --1.46 &  2 \\     
  NGC 6093 &  M80   & --1.75        &  1 &    NGC 6656 & M22   & --1.70 &  1 \\     
  NGC 6205 &  M13   & --1.58        &  3 &    NGC 6715 & M54   & --1.44 &  4 \\     
  NGC 6218 &  M12   & --1.33        &  1 &    NGC 6749 &       & --1.62 &  1 \\     
  NGC 6229 &        & --1.43        &  2 &    NGC 6752 &       & --1.55 &  1 \\     
  NGC 6254 &  M10   & --1.57        &  2 &    NGC 6779 & M56   & --2.00 &  2 \\     
  NGC 6256 &        & --0.62        &  1 &    NGC 7078 & M15   & --2.33 &  3 \\     
  NGC 6266 &  M62   & --1.18        &  3 &    NGC 7089 & M2    & --1.66 &  4 \\     
  NGC 6273 &  M19   & --1.76        &  4 &       Pal 3 &       & --1.67 &  1 \\     
  NGC 6284 &        & --1.31        &  2 &    Terzan 1 &       & --1.29 &  1 \\     
  NGC 6325 &        & --0.90        &  2 &             &       &        &    \\     
\hline     
\end{tabular}     
\tablefoot{     
\scriptsize     
Source of the metallicities:      
\citet{harris96}, transformed to the \citet{carretta09}     
scale. For V1, V29 and V48 in \wcen we adopted metallicities from     
\citet{gonzalez1997b} converted to the current metallicity scale,     
based on a solar iron abundance by number of $\log{\epsilon Fe}$=7.54 dex \citep{gratton2003}.     
For the other TIICs in \wcen (all BLHs), we adopted --1.61 dex, which is the      
metallicity of V48 (the only BLH with a spectroscopic estimate     
of the metallicity). This assumption agrees with the      
metallicity distribution found by \citet{johnson10} and      
\citet{magurno2019} in \wcen, moreover, it is well within the      
standard deviation of the metallicity distribution of this cluster.     
For NGC~6325, we adopted the iron abundance derived by      
\citet{minniti1995}. We did not transform it to our      
scale because $\log{\epsilon Fe}$ is not available,      
and the standard deviation is much larger (0.30 dex)      
than the correction.}     
\end{table}     
     
%%%%%%%%%%%%%%%%%%%%%%%%%%%%%%%%%%%%%%%%%%%%%%%%%%%%%%%%%%%%%%%%%%%%%%%%5  
% Table A-2 
%%%%%%%%%%%%%%%%%%%%%%%%%%%%%%%%%%%%%%%%%%%%%%%%%%%%%%%%%%%%%%%%%%%%%%%%5  
\begin{table}     
\scriptsize     
\caption{Iron abundances for Galactic field TIICs.}     
\label{tab:field_met}     
\centering     
\begin{tabular}{lrclrc}     
\hline     
\hline     
Name &  [Fe/H] & source & Name &  [Fe/H] & source \\     
\hline     
AL CrA         & --0.42 &  0 & SW Tau         &   0.15 &  1 \\     
AL Vir         & --0.48 &  1 & SZ Mon         & --0.52 &  1 \\     
AP Her         & --0.84 &  1 & TX Del         &   0.02 &  1 \\     
AU Peg         & --0.28 &  1 & VY Pyx         & --0.49 &  1 \\     
BL Her         & --0.22 &  1 & VZ Aql         &   0.34 &  2 \\     
BO Tel         & --0.51 &  0 & V439 Oph       & --0.34 &  2 \\     
BX Del         & --0.26 &  1 & V446 Sco       & --1.12 &  0 \\     
CC Lyr         & --3.98 &  1 & V449 CrA       & --0.51 &  0 \\     
CO Pup         & --0.73 &  1 & V478 Oph       & --0.86 &  0 \\     
CQ Sco         & --1.46 &  0 & V553 Cen       &   0.03 &  4 \\     
CS Cas         & --0.60 &  0 & V554 Oph       & --1.20 &  0 \\     
DD Vel         & --0.45 &  3 & V709 Sco       & --0.77 &  0 \\     
EP Lyr         & --1.80 &  8 & V745 Oph       & --0.74 &  2 \\     
HQ Car         & --0.29 &  3 & V802 Sgr       & --0.60 &  0 \\     
IX Cas         & --0.57 &  1 & V971 Aql       & --0.34 &  2 \\     
$k$ Pav     &   0.06 &  2 & V1004 Sgr      & --0.42 &  0 \\     
KQ CrA         & --0.86 &  0 & V1185 Sgr      & --0.68 &  0 \\     
MR Ara         & --1.12 &  0 & V1189 Sgr      & --1.20 &  0 \\     
MZ Cyg         & --0.27 &  1 & V1290 Sgr      & --1.29 &  0 \\     
NW Lyr         & --0.14 &  2 & V1303 Sgr      & --0.94 &  0 \\     
QQ Per         & --0.70 &  6 & V1304 Sgr      &   0.01 &  0 \\     
PP Aql         & --0.25 &  0 & V1711 Sgr      & --1.26 &  1 \\     
RR Mic         & --1.46 &  0 & W Vir          & --1.06 &  1 \\     
RX Lib         & --1.04 &  1 & XX Vir         & --1.61 &  2 \\     
ST Pup         & --1.47 &  5 & YZ Vir         & --1.03 &  0 \\     
\hline     
\end{tabular}     
\tablefoot{Sources of the metallicities:       
0: \citet{harris_wallerstein1984}     
1: \citet{maas2007}     
2: \citet{wallerstein_farrell2018}     
3: \citet{lemasle15}     
4: \citet{wallerstein1996}     
5: \citet{gonzalez96}     
6: \citet{wallerstein2008}     
7: \citet{wallerstein2000}     
8: \citet{gonzalez1997a}.   
Note that we included in this table CC Lyr even tough  
its metal abundance is quite suspicious.  
}     
\end{table}     
     
Individual high-resolution (HR) spectroscopic      
abundances are available for 29 field TIICs      
\citep{wallerstein1996,gonzalez96,gonzalez1997a,wallerstein2000,maas2007,wallerstein2008,lemasle15,wallerstein_farrell2018}      
and 3 cluster TIICs \citep{gonzalez1994b}.     
We have rescaled all the [Fe/H] values by adopting a solar iron abundance      
by number of $\log{\epsilon Fe}$=7.54 dex \citep{gratton2003}.     
\citet{harris_wallerstein1984} provided metallicities      
([A/H] in their Table 2) for 50 variable stars from       
low-resolution (LR) spectra. In their sample 38 objects are TIICs, and 17 are in      
common with our HR metallicity sample.      
Based on the stars in common, we found an empirical      
relation between \citet{harris_wallerstein1984} metallicity      
scale and the homogeneous HR iron abundances ($[Fe/H]_{HR}=0.864*[A/H]_{LR}-0.251$).      
We have adopted this relation to transform the LR     
metallicities of the remaining 21 TIICs to the HR      
metallicity scale. We ended up with a sample of 50 field TIICs with      
metal abundance determinations.       
     
The metallicity distribution of cluster and field TIICs was investigated by \cite{harris81},      
using Washington photometry. The resulting distributions were quite      
similar (see his Figure~8) to the current ones even if they were based on a photometric index.      
In particular, he suggested that only a minor fraction of TIICs      
appears to be a true Halo population, indeed, a significant fraction of them appeared to be      
a transitional population between the Halo and the disk.      
%     
     
%______________________________________________________________________________     
\begin{figure}[!htbp]     
\centering     
\includegraphics[width=8.5cm]{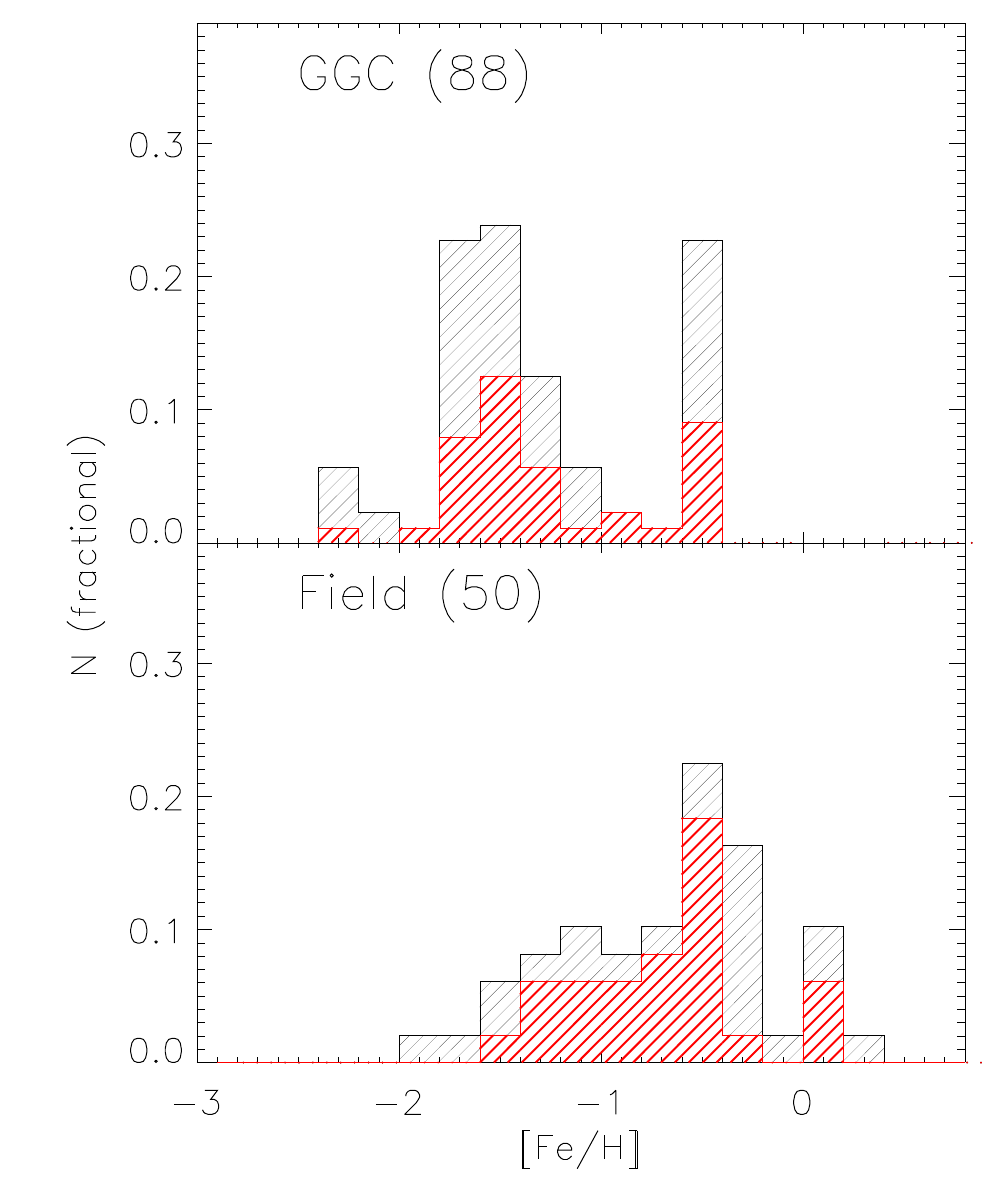}     
\caption{     
Top: Metallicity distribution of cluster TIICs. The source catalog is     
on the database by \citet{clement01}. The iron abundances are based on the     
metallicity scale by \citet{carretta09}.     
Bottom: Same as the top panel, but for Galactic field TIICs. The iron abundances are     
based on high- and low-resolution spectra     
The red histogram display the metallicity distribution of WVs only (see text for more details).     
}     
\label{fig:metallicity}     
\end{figure}     
% figura generata con IDLWorkspace/Default/Temp/temp_181213.pro     
     
Data plotted in the bottom panel of Fig.~\ref{fig:metallicity} show      
that field TIICs cover a broad range in metallicity, from [Fe/H]$\sim$--2      
to [Fe/H]$\sim$0.3. The distribution appears to be similar to     
field RRLs \citep{magurno2018,fabrizio2019}, suggesting a similar evolutionary channel.         
Moreover, BLHs, WVs and RVTs appear to have very similar metallicity distributions.      
Recent spectroscopic investigations based on 10 objects suggested that       
short-period (P$<$3 days) TIICs are separated into metal-rich ([Fe/H]$\gtrsim$--0.5 dex)      
labeled as BLHs and metal-poor (--2.0$\lesssim$[Fe/H]$\lesssim$--1.5), labeled as      
UY Eri (UYE) stars \citep{kovtyukh2018a}. Moreover, it was also suggested that field      
WVs are more metal-poor than --0.3 dex, while the current sample is suggesting a      
metallicity distribution approaching, within the errors, either solar or super solar      
iron abundances.     
     
The elemental abundances of TIICs have a long-standing record. Dating back to more      
than half century ago, \citet{rodgers1968} and later on \citet{luck1989}     
found that TIICs are deficient in s-process elements. More recently,      
\citet{maas2007} found solid evidence of a contamination with 3$\alpha$ and CN-cycling products      
in the CNO abundances of field BLHs and WVs. Moreover, they found a clear Na overabundance in      
BLHs, but not in WVs. There is also evidence that [Ca/Fe] and [Ti/Fe] are under-abundant in      
WVs thus suggesting the possible presence of a gas-dust separation \citep{wallerstein2000}.

%______________________________________________________________________________     
\begin{figure}[!htbp]     
\centering     
\includegraphics[width=8.5cm]{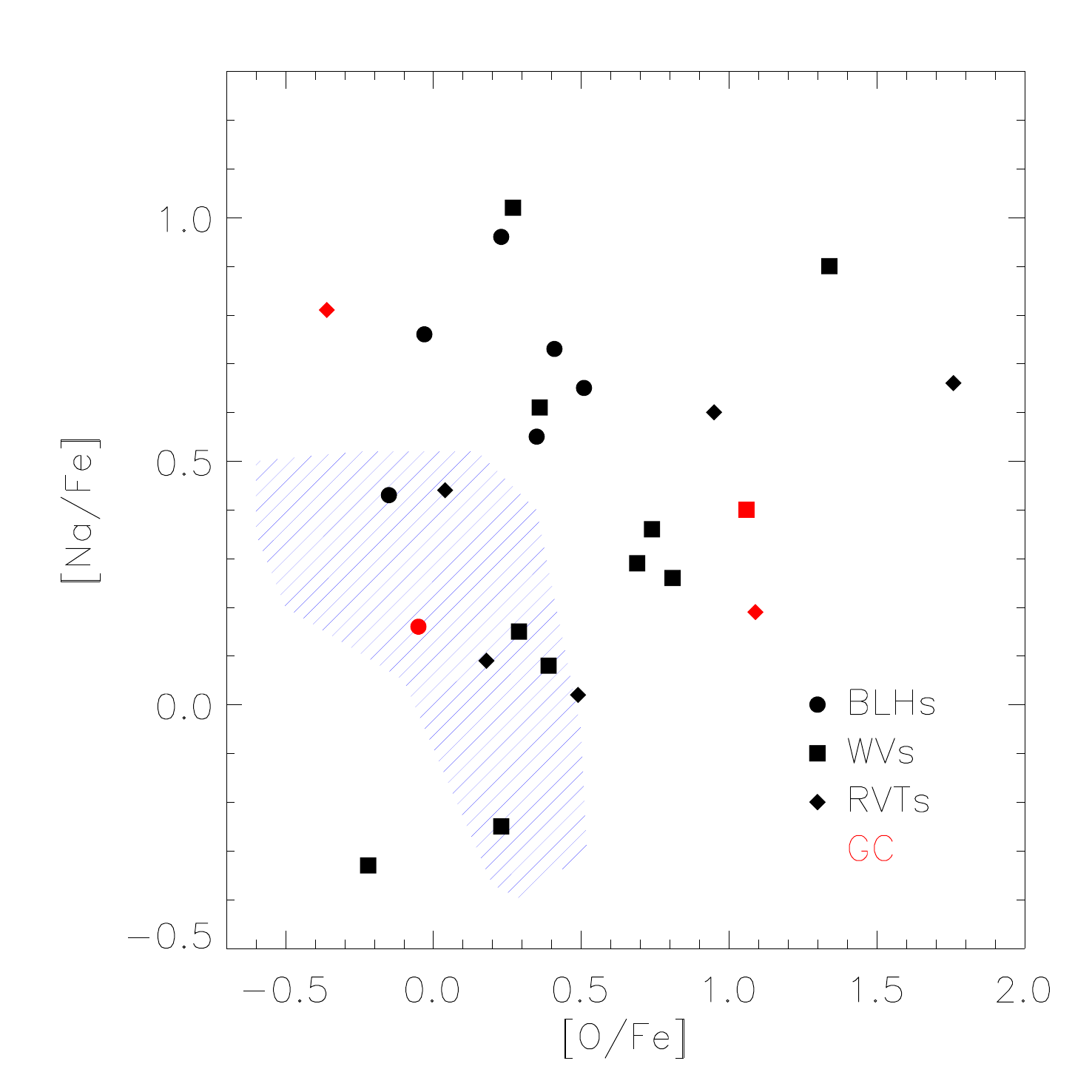}     
\caption{Na-O anticorrelation for field (black) and cluster (red) TIICs.   
The three different sub-groups are marked with different symbols (see labels).     
The blue hatched area marks the area covered by cluster stars   
showing a well defined anti-correlation between Na and O \citep{gratton04}.   
}     
\label{fig:Na_O_anti}     
\end{figure}     
% figura generata con temp_200403   
     
To investigate in more detail TIIC chemical peculiarities, and in particular  
to assess possible differences between field and cluster TIICs, Fig.~A.2  
shows the measured [Na/Fe] versus [O/Fe] abundance ratios for field and cluster objects.   
A glance at the data shows that Na and O   
overabundances in field (black symbols) and cluster (red symbols) TIICs span    
Oven one dex. There is a marginal evidence for WVs (squares) and RVTs   
(diamonds) to be more overabundant in O compared to BLHs (circles),   
while BLHs appear to be more overabundant in Na. However, the sample of TIICs   
with accurate measurements (two dozen) is too limited to reach firm conclusions.       
We also note that roughly one third of the current sample overlaps   
with values typical of cluster stars showing a well-defined anti-correlation   
between Na and O \citep{gratton04}. Note that the cluster   
TIICs located in this area is V48 in $\omega$ Cen, and more quantitative   
discussions concerning the possible difference between first and second generation   
stars are hampered by the fact that accurate abundances are only available for   
two out of the seven TIICs present in this cluster.         
  
The above chemical peculiarities for TIICs appear more in context if we take into account      
the evolutionary properties of their progenitors. Dating back to the seminal    
theoretical investigations by     
\citet{giannone77} and the empirical works by \citet{kudritzki86} and \citet{behr03},       
it has become clear that hot and extreme HB stars are affected by gravitational settling      
and radiative levitation of heavier elements \citep{cassisi13}. In addition,     
there is the possibility of having hot helium flashers,     
stars that do not experience the core helium flash at the tip      
of the red giant branch, and whose surface abundances are altered   
by mixing processes during the flash \citep{castellani93,d'cruz96, brown00, cassisi03}.  
Recent spectroscopic investigations of extreme HB stars have revealed      
well defined patterns in their surface chemical composition, that has been altered   
by chemical element transport processes \citep{moehler11}.    
Moreover, in a recent investigation \citet{latour14} found a well defined    
Helium-Carbon correlation among extreme HB stars in $\omega$ Cen suggesting    
the occurrence of diffusion mechanisms  \citep{millerbertolami08}.     

It is clear that cluster TIICs can play a crucial role in addressing some of the      
current open problems affecting the origin and evolution of TIICs. Indeed for    
these objects, we have detailed information concerning the age and the chemical    
composition of their progenitors.     
\end{appendix}  
  
     
%%%%%%%%%%%%%%%%%%%%%%%%%%%%%%%%%%%%%%%%%%%%%%%%%%     

% Don't change these lines     
% \bsp	% typesetting comment     
% \label{lastpage}     
\end{document}